%% file: main.tex
\documentclass[journal]{IEEEtran}  %

\usepackage[utf8]{inputenc}
\usepackage{amsmath,amssymb,amsfonts} %
\usepackage{graphicx} %
\usepackage{subcaption} %
\usepackage{cite} %
\usepackage{booktabs} %
\usepackage{multirow} %
\usepackage{algpseudocode} %
\usepackage{algorithm} %
\usepackage{xcolor} %
\usepackage{textcomp} %
\usepackage{tikz} %
\usepackage{adjustbox} %
\usepackage{rotating} %
\usepackage{pifont} %

\usepackage{tikz}
\usepackage{subcaption}
\usepackage{array}
\usepackage{caption}
\usepackage{tabularx}
\usepackage{makecell}
\usepackage{lscape}
\usepackage{colortbl}   %
\usepackage{setspace}
\usepackage{bm}
\usepackage{enumitem}
\usepackage{xcolor}
\usepackage{titlesec}
\usepackage{mdframed}

\usepackage{pgfplots}
\usepgfplotslibrary{groupplots}
\pgfplotsset{compat=1.18}
\newcommand{\revised}[1]{{\color{black}#1}}

\begin{document}

\bstctlcite{IEEEexample:BSTcontrol}
\title{VitaLLM: A Versatile, Ultra-Compact Ternary LLM Accelerator with Dependency-Aware Scheduling}

\author{\IEEEauthorblockN{Zi-Wei Lin, and Tian-Sheuan Chang, \textit{Senior Member, IEEE}}
\thanks{This work has been accepted to be published in IEEE Transactions on Circuits and Systems for Artificial Intelligence.This work was supported by the National Science and Technology Council, Taiwan, under Grant  113-2221-E-A49-078-MY3, 113-2640-E-A49-005, 113-2221-E-A49-078-MY3 and 114-2640-E-A49-011. The authors are with the Institute of Electronics, National Yang Ming Chiao Tung University, Taiwan. (e-mail: ziweii0908.ee13@nycu.edu.tw; tschang@nycu.edu.tw). }%
\thanks{Manuscript received XXXX XX, 2025; revised XXXX XX, XXXX.}
}
\maketitle

\input{abs/abstract_en}

\input{chapters/1Introduction}
\input{chapters/2HardwareImplementation}
\input{chapters/3SystemIntegrationandScheduling}
\input{chapters/4ExperimentalEvaluation}
\input{chapters/5ExtendedDesign}
\input{chapters/6Conclusion}

\bibliographystyle{IEEEtran}

\bibliography{bib/ieeeBSTcontrol,bib/thesis}

\begin{IEEEbiography}[{\includegraphics[width=1in,height=1.25in,clip,keepaspectratio]{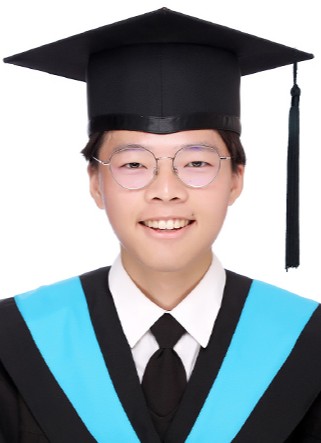}}]{Zi-Wei Lin}
received the B.S. and M.S. degrees in electronic engineering from the National Yang Ming Chiao Tung University (NYCU), Hsinchu, Taiwan, in 2024 and 2025, respectively.
His research interests include digital integrated circuit design, computer architecture, and hardware accelerators for artificial intelligence.
\end{IEEEbiography}

\begin{IEEEbiography}[{\includegraphics[width=1in,height=1.25in,clip,keepaspectratio]{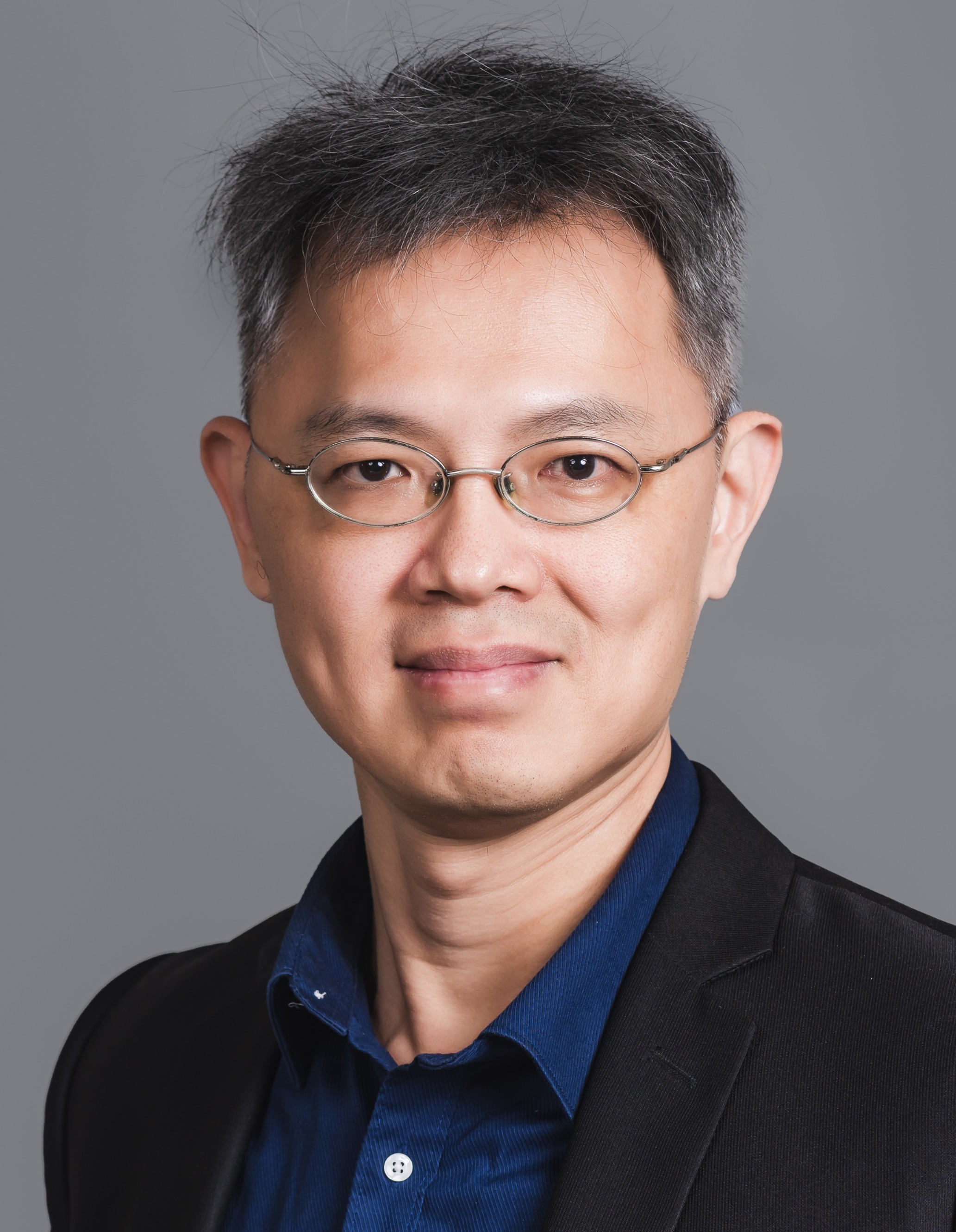}}]{Tian-Sheuan Chang}
	(S’93–M’06–SM’07)
	received the B.S., M.S., and Ph.D. degrees in electronic engineering from National Chiao-Tung University (NCTU), Hsinchu, Taiwan, in 1993, 1995, and 1999, respectively. 
	
	From 2000 to 2004, he was a Deputy Manager with Global Unichip Corporation, Hsinchu, Taiwan. In 2004, he joined the Department of Electronics Engineering, NCTU (as National Yang Ming Chiao Tung University (NYCU) in 2021), where he is currently a Professor. In 2009, he was a visiting scholar in IMEC, Belgium. His current research interests include system-on-a-chip design, VLSI signal processing, and computer architecture.
	
	Dr. Chang has received the Excellent Young Electrical Engineer from Chinese Institute of Electrical Engineering in 2007, and the Outstanding Young Scholar from Taiwan IC Design Society in 2010. He has been actively involved in many international conferences as an organizing committee or technical program committee member.
\end{IEEEbiography}
\end{document}

%% file: abs/abstract_en.tex
\begin{abstract}
Deploying Large Language Models (LLMs) on resource-constrained edge devices faces critical bottlenecks in memory bandwidth and power consumption. While ternary quantization (e.g., BitNet b1.58) significantly reduces model size, its direct deployment on general-purpose hardware is hindered by workload imbalance, bandwidth-bound decoding, and strict data dependencies. To address these challenges, we propose \textbf{VitaLLM}, a hardware-software co-designed accelerator tailored for efficient ternary LLM inference. We introduce a heterogeneous \textbf{Dual-Core Compute Strategy} that synergizes specialized TINT-Cores for massive ternary projections with a unified BoothFlex-Core for mixed-precision attention, ensuring high utilization across both compute-bound prefill and bandwidth-bound decode stages. Furthermore, we develop a \textbf{Leading One Prediction (LOP)} mechanism to prune redundant Key-Value (KV) cache fetches and a \textbf{Dependency-Aware Scheduling} framework to hide the latency of nonlinear operations. Implemented in TSMC 16nm technology, VitaLLM achieves a decoding throughput of 70.70 tokens/s within an ultra-compact area of 0.223 mm$^2$ and a power consumption of 65.97 mW. The design delivers a superior Figure of Merit (FOM) of 17.4 TOPS/mm$^2$/W, significantly outperforming state-of-the-art accelerators. Finally, we explore an extended bit-serial design (BoothFlex-BS) to demonstrate the architecture's adaptability for precision-agile inference.
\end{abstract}

%% file: chapters/1Introduction.tex
\section{Introduction}
\label{sec:introduction}

The rapid evolution of Large Language Models (LLMs) has fundamentally transformed natural language processing. However, deploying these billion-parameter models is typically confined to high-performance cloud servers. As the demand for privacy-preserving, low-latency intelligence on resource-constrained edge devices grows, three critical hardware bottlenecks must be addressed: limited on-chip memory capacity, restricted DRAM bandwidth, and strict power budgets~\cite{mobilellm2024}.

To address these inefficiencies, ternary quantization (e.g., BitNet b1.58~\cite{bitnet158, bitnet2b4t}) has emerged as a promising solution. By constraining weights to $\{-1, 0, +1\}$ while maintaining 8-bit activations, BitNet b1.58 drastically reduces the model footprint and theoretically replaces expensive floating-point multiplications with simple additions. However, realizing this potential requires addressing the critical architectural mismatches—namely, the highly unbalanced workload between ternary projections and high-precision attention, the bandwidth-bound autoregressive decoding, and the strict data dependencies in nonlinear operations.

Prior efforts to deploy efficient LLMs on the edge have met with mixed success. General-purpose processors (e.g., mobile GPUs and NPUs), while optimized for dense INT8/FP16 arithmetic, are fundamentally inefficient for ternary models. As noted in recent studies~\cite{bitnet_cpp, survey_hw_llm}, executing 1.58-bit operations on fixed 8-bit or 16-bit datapaths results in significant resource underutilization. Furthermore, these architectures lack the specialized decoding logic required to handle packed ternary weights efficiently, often negating the bandwidth benefits of quantization due to software unpacking overheads~\cite{bitnet_cpp}.

To overcome these limitations, specialized accelerators have been proposed. In the ASIC domain, \textbf{Slim-Llama}~\cite{slimllama2025} introduces output-reuse schemes, but their efficiency diminishes during the bandwidth-bound decode stage where weight reuse is minimal. In the FPGA domain, designs like \textbf{TerEffic}~\cite{tereffic} and \textbf{TeLLMe}~\cite{tellme2025} optimize storage and scheduling but typically rely on massive on-chip memory resources (e.g., URAM) that are unavailable in cost-sensitive mobile SoCs.

To address these gaps and enable efficient edge inference for BitNet b1.58 while adhering to strict area constraints, we propose \textbf{VitaLLM}, an ultra-compact accelerator featuring:
\begin{itemize}
    \item \textbf{Dual-Core Compute Strategy:} A heterogeneous architecture featuring specialized TINT-Cores for massive ternary projections and a unified BoothFlex-Core for mixed-precision tasks. This decoupling ensures high utilization across both prefill and decode stages.
    \item \textbf{Leading One Prediction (LOP):} A unified predictor that utilizes the leading `1' bit position to identify critical tokens. This mechanism prunes redundant KV cache fetches, significantly mitigating the bandwidth bottleneck in the decode stage.
    \item \textbf{Dependency-Aware Scheduling:} A framework comprising Head-Level Pipelining, Two-stage Nonlinear Operations and Q-Friendly Two-Level Scheduling. These techniques effectively hide the latency of global reductions and quantization dependencies.
\end{itemize}

Implemented in TSMC 16nm technology, VitaLLM achieves 70.70 tokens/s in a compact 0.223 $\text{ mm}^2$ area. The following sections detail the hardware implementation and system integration.

The remainder of this paper is organized as follows. Section~\ref{sec:hardware_implementation} details the hardware architecture, including the heterogeneous core designs and memory hierarchy. Section~\ref{sec:system_integration} presents the system integration strategies, focusing on the LOP mechanism and dependency-aware scheduling. Section~\ref{sec:experimental_evaluation} provides the experimental results and comparisons with state-of-the-art works. Section~\ref{sec:extended_design} explores the extended precision-agile design, and Section~\ref{sec:conclusion} concludes the paper.

%% file: chapters/2HardwareImplementation.tex
\section{Hardware Implementation}
\label{sec:hardware_implementation}

To address the stringent constraints of edge LLM deployment, VitaLLM adopts a hardware-software co-design approach. In this section, we first detail the operational principles of the target model, analyze the specific design challenges derived from these characteristics, and then present the proposed heterogeneous architecture.

\subsection{Preliminaries: BitNet b1.58 Architecture}
\label{subsec:bitnet_background}

\begin{figure}[t]
    \centering
    \includegraphics[width=1.0\linewidth]{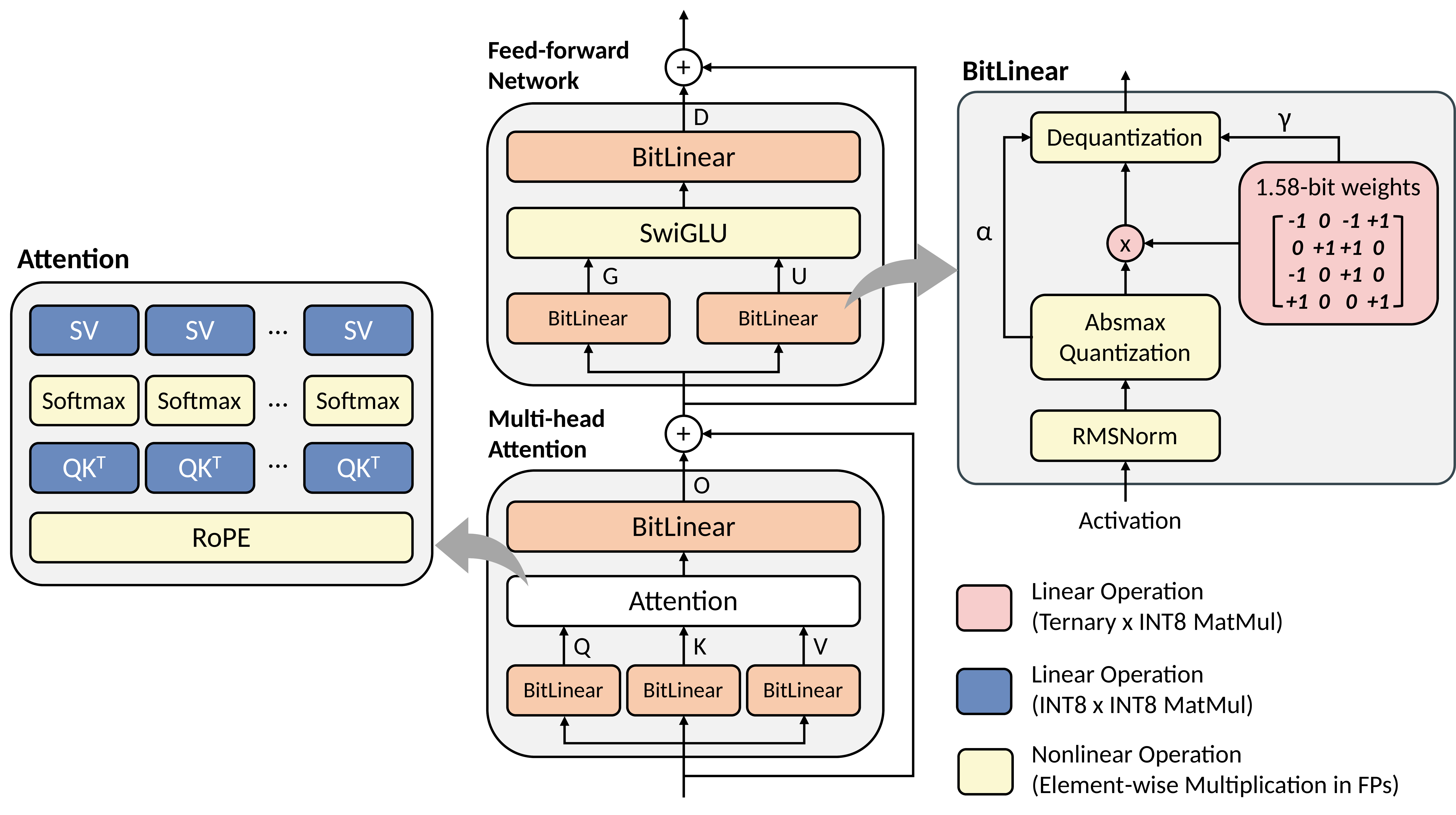}
    \caption{Overview of the BitNet b1.58 architecture. Conventional linear projection layers are replaced by BitLinear modules using Ternary$\times$INT8 arithmetic.}
    \label{fig:bitnet_architecture}
\end{figure}

BitNet b1.58~\cite{bitnet158, bitnet2b4t} represents a significant breakthrough in 1-bit LLMs, trained from scratch to preserve competitive accuracy. As illustrated in Fig.~\ref{fig:bitnet_architecture}, the model replaces standard linear projection layers with \textbf{BitLinear} modules.

In these modules, weights $W$ are constrained to ternary values $\{-1, 0, +1\}$, while input activations $x$ are quantized to 8-bit integers. Since $w \in \{-1, 0, +1\}$, the computationally expensive floating-point MAC operations are effectively replaced by simple additions and subtractions. While this design significantly lowers arithmetic intensity, it introduces unique dataflow requirements that differ from conventional INT8 models.

\subsection{The Analysis of Design Challenges }
\label{subsec:design_challenges}

Straightforward deployment of ternary LLMs on edge hardware faces three specific architectural mismatches:

\subsubsection{Workload Imbalance and Heterogeneity}
The computational workload in BitNet b1.58 is highly unbalanced. Ternary weight projections dominate the operation count (approximately 94.68\% \revised{in BitNet b1.58 3B}), while high-precision INT8$\times$INT8 attention computations account for only a small fraction (5.32\%). A homogeneous architecture designed for a single precision would inevitably suffer from low utilization. Furthermore, the inference alternates between a compute-bound \textbf{Prefill stage} (parallel matrix-matrix processing) and a bandwidth-bound \textbf{Decode stage} (sequential vector-matrix generation), imposing conflicting requirements on hardware parallelism.

\subsubsection{Memory Bandwidth Bottleneck}
In the autoregressive decode stage, the lack of weight reuse forces the system to fetch the entire weight matrix for every single token, making performance strictly constrained by DRAM bandwidth. This bottleneck is exacerbated by the attention mechanism, which must fetch the Key (K) and Value (V) cache for all previous tokens at every step. Standard implementations fetch the entire cache regardless of relevance, resulting in massive redundant data movement that saturates the memory interface.

\subsubsection{Complex Data Dependencies}
BitNet b1.58 introduces strict dependencies in nonlinear operations (Softmax, RMSNorm) and quantization. These operations require global statistics (e.g., global maximum or sum of squares) from the entire vector before processing individual elements. This requirement creates "stop-and-wait" barriers that prevent efficient tile-level pipelining, forcing the hardware to stall and wait for full-vector completion.

Driven by these challenges, the architecture of VitaLLM targets the deployment of the BitNet b1.58 3B model. The design integrates heterogeneous computing cores to address the workload imbalance and specialized buffering strategies to mitigate bandwidth limitations.

\subsection{Overall Architecture}
\label{subsec:overall_architecture}

The architecture of VitaLLM targets the deployment of the \textbf{BitNet b1.58 3B} model on resource-constrained edge devices. To ensure real-time interactivity, the design targets a prefill latency of less than \textbf{1.0 second} and a decoding throughput exceeding \textbf{20 tokens/s}. 

The system operates under mobile bandwidth limitations, modeled after the \textbf{LPDDR5-9600} standard with a 2-channel configuration (theoretical peak of \textbf{76.8 GB/s}~\cite{liang2024computing}). A critical design challenge is balancing on-chip computing resources against this bandwidth to maintain high utilization across both the compute-bound prefill stage and the bandwidth-bound decode stage.

\begin{figure}[t]
    \centering
    \includegraphics[width=0.9\linewidth]{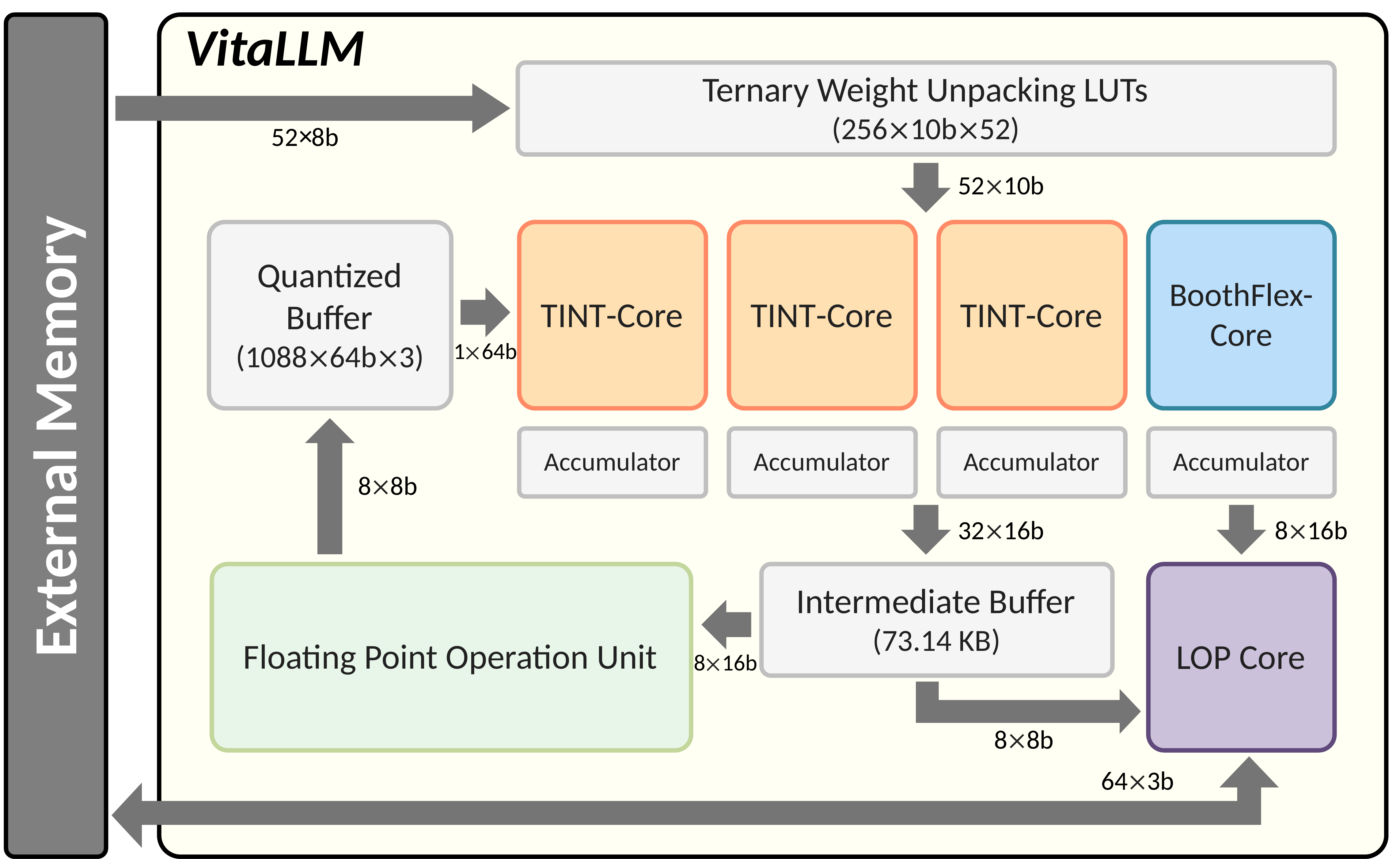}
    \caption{Top-level block diagram of the VitaLLM accelerator. The system integrates heterogeneous computing cores (TINT-Cores and BoothFlex-Core), a Leading One Prediction (LOP) Core for sparse attention, and a hierarchical on-chip memory system.}
    \label{fig:overall_architecture}
\end{figure}
\begin{figure}[t]
    \centering
    \includegraphics[width=0.9\linewidth]{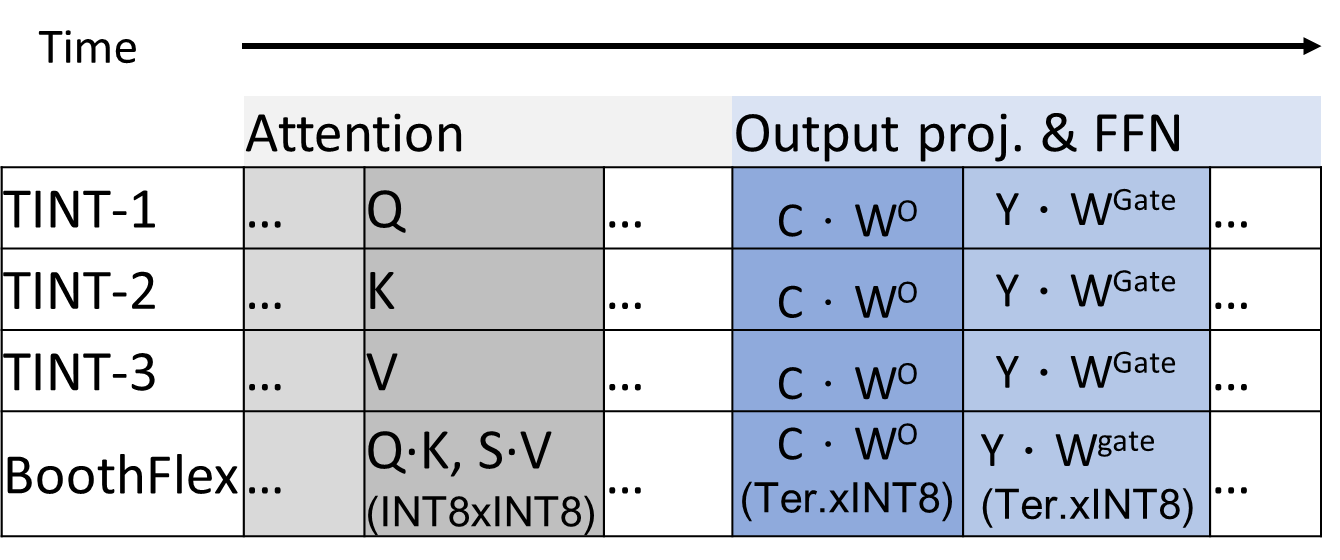}
    \caption{Schedule of the TINT-Cores and BoothFlex-Core.}
    \label{fig:overall_core_schedule}
\end{figure}

\revised{

Fig.~\ref{fig:overall_architecture} shows the proposed heterogeneous architecture with three distinct computing units:
\begin{enumerate}
    \item \textbf{TINT-Cores:} A lightweight compute cluster specialized for the dominant Ternary$\times$INT8 matrix multiplications in BitNet linear projections, providing high compute density and energy efficiency for the main projection workload.
    \item \textbf{BoothFlex-Core:} A shared mixed-precision engine that executes latency-critical INT8$\times$INT8 attention operations and is reconfigured to support Ternary$\times$INT8 projections during WO and FFN execution, thereby assisting the TINT-Cores and improving overall hardware utilization, as shown in Fig.~\ref{fig:overall_core_schedule}.
    \item \textbf{Leading One Prediction (LOP) Core:} A dedicated sparse-attention unit that predicts the most relevant KV entries to reduce redundant KV-cache accesses and unnecessary attention computation during autoregressive decoding.
\end{enumerate}
This heterogeneous architecture is due to the workload heterogeneity as in Fig.~\ref{fig:bitnet_architecture}. However, the resulted core design (TINT-core for ternary$\times$INT8 and Boothflex-core for INT8$\times$INT8) based on this workload  will lead to the idle Boothflex-core during WO and FFN execution. Thus, to avoid this while reduce latency, we reconfigure Boothflex-core for ternary$\times$INT8 as well to assist TINT-cores during these phases. Scaling TINT-cores to support INT8$\times$INT8 will require considerable datapath replication and reduction logic to recover throughput as Boothflex-core, which significantly diminishes the original area benefit. 
Overall, the proposed heterogeneous organization balances specialization and utilization by assigning the dominant ternary workload to the lightweight TINT-Cores, while reusing the BoothFlex-Core for both indispensable attention computation and opportunistic assistance in WO/FFN execution.

}

To bridge the gap between compute density and memory bandwidth, VitaLLM employs a specialized memory hierarchy comprising a \textbf{Quantized Buffer} for activation vectors and an \textbf{Intermediate Buffer} for nonlinear processing results. To maximize effective bandwidth, \textbf{Ternary Weight Unpacking Look-Up Tables (LUTs)} are deployed at the memory interface to decode compressed weights on-the-fly. Finally, a \textbf{Floating-Point Operation Unit} handles element-wise and nonlinear operations (e.g., RMSNorm, Softmax) to complete the end-to-end inference pipeline.

\subsection{TINT-Core Design}
\label{subsec:tint_core}

The TINT-Core is specialized for accelerating Ternary$\times$INT8 matrix multiplications, which constitute the main workload. By exploiting the limited value set of ternary weights, general-purpose multipliers are replaced with efficient selection logic.

\subsubsection{Microarchitecture}
\begin{figure}[t]
    \centering
    \includegraphics[width=0.8\linewidth]{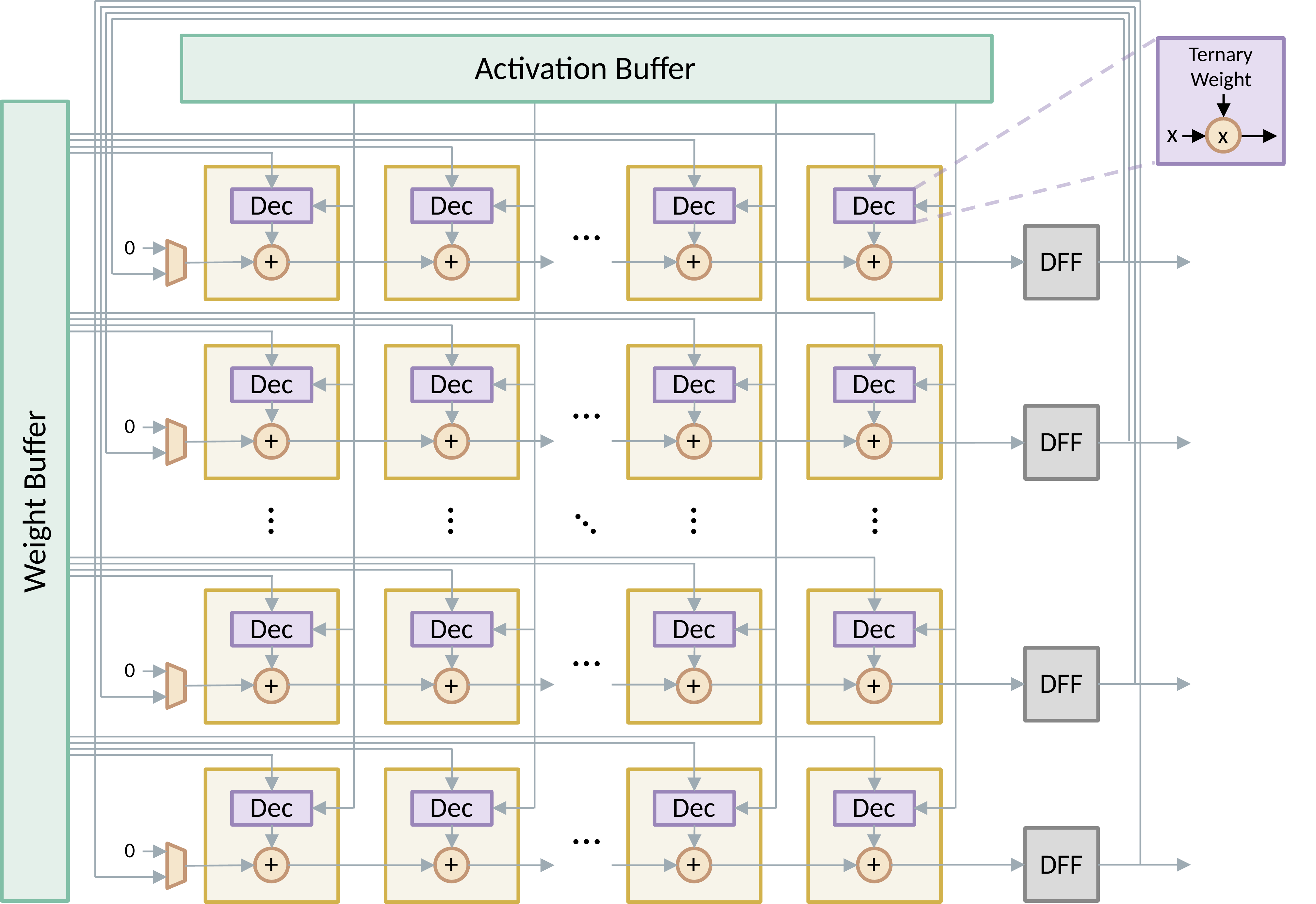}
    \caption{Architecture of the TINT-Core featuring an $8\times8$ PE array. \revised{Multiplier-free logic uses local selectors for partial sum updates.} }
    \label{fig:tint_core_architecture}
\end{figure}

Fig.~\ref{fig:tint_core_architecture} depicts the TINT-Core, centered around an 8$\times$8 Processing Element (PE) array. The multiplication of an activation $a$ by a weight $w$ is simplified to a selection operation:
\begin{equation}
    y \leftarrow y + \text{sel}(w, a), \quad \text{where } \text{sel}(w, a) \in \{+a, 0, -a\}
\end{equation}
Each PE employs a lightweight selector controlled by a local ternary decoder. This design significantly reduces silicon area compared to standard MAC units. The array sustains 64 operations per cycle, with activations broadcast across columns and unique weights unicast to each PE.

\subsubsection{Byte-Level Ternary Weight Packing}
To mitigate memory bandwidth bottlenecks, we adopt the dense weight packing strategy proposed in TerEffic~\cite{tereffic}. This method packs five ternary weights into a single 8-bit byte, achieving an effective density of 1.6 bits per weight. This yields an approximately 20\% reduction in memory traffic compared to naive 2-bit encoding. A hardware unpacking unit, implemented as a precomputed lookup table (LUT), decodes these packed bytes on-the-fly between the weight buffer and the PE array.

\subsubsection{Dataflow}
\label{sec:TINT_dataflow}
\begin{figure}[t]
    \centering
    \begin{subfigure}[b]{0.23\textwidth}
        \centering
        \includegraphics[width=\textwidth]{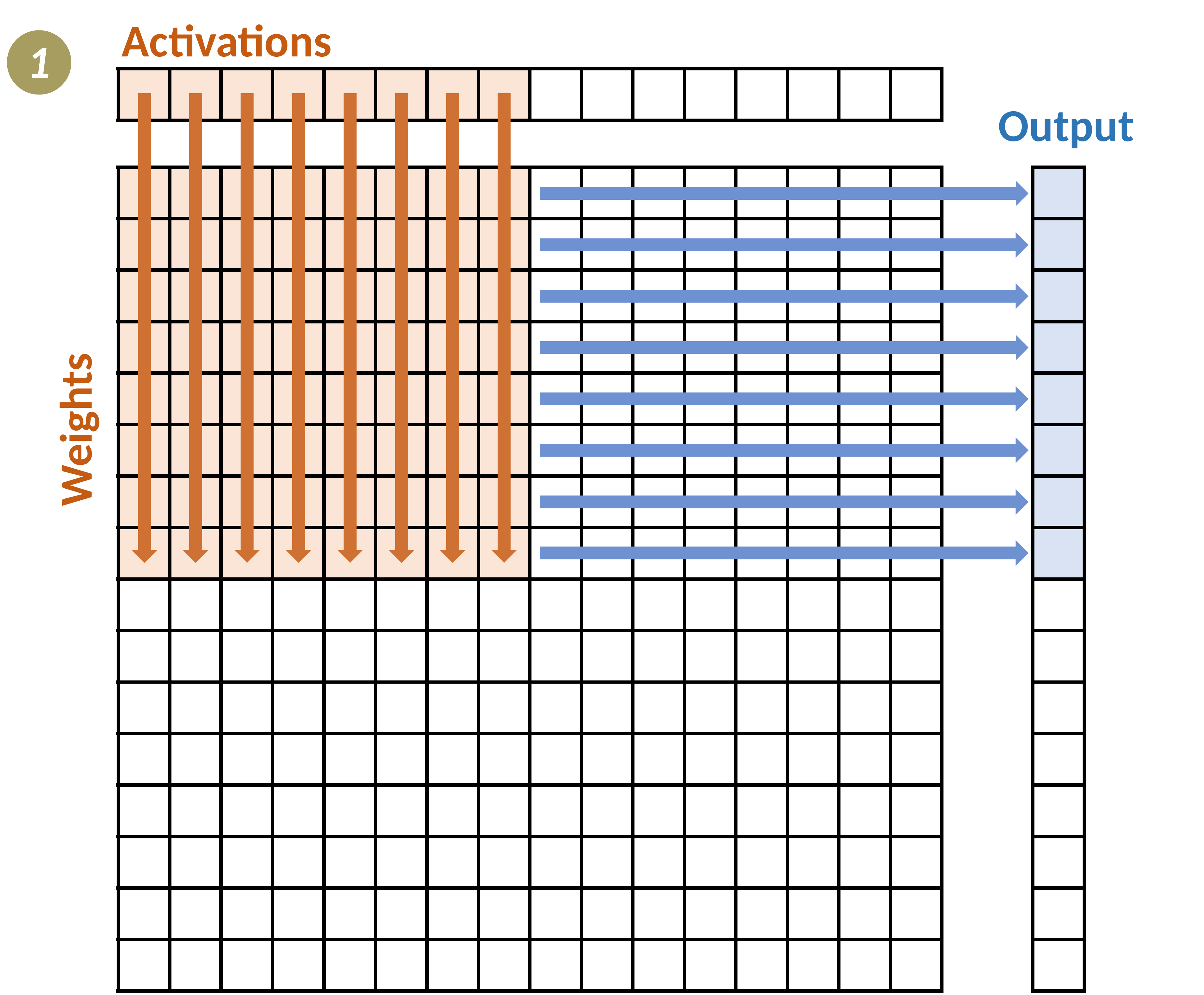}
        \label{fig:sub1}
    \end{subfigure}
    \begin{subfigure}[b]{0.23\textwidth}
        \centering
        \includegraphics[width=\textwidth]{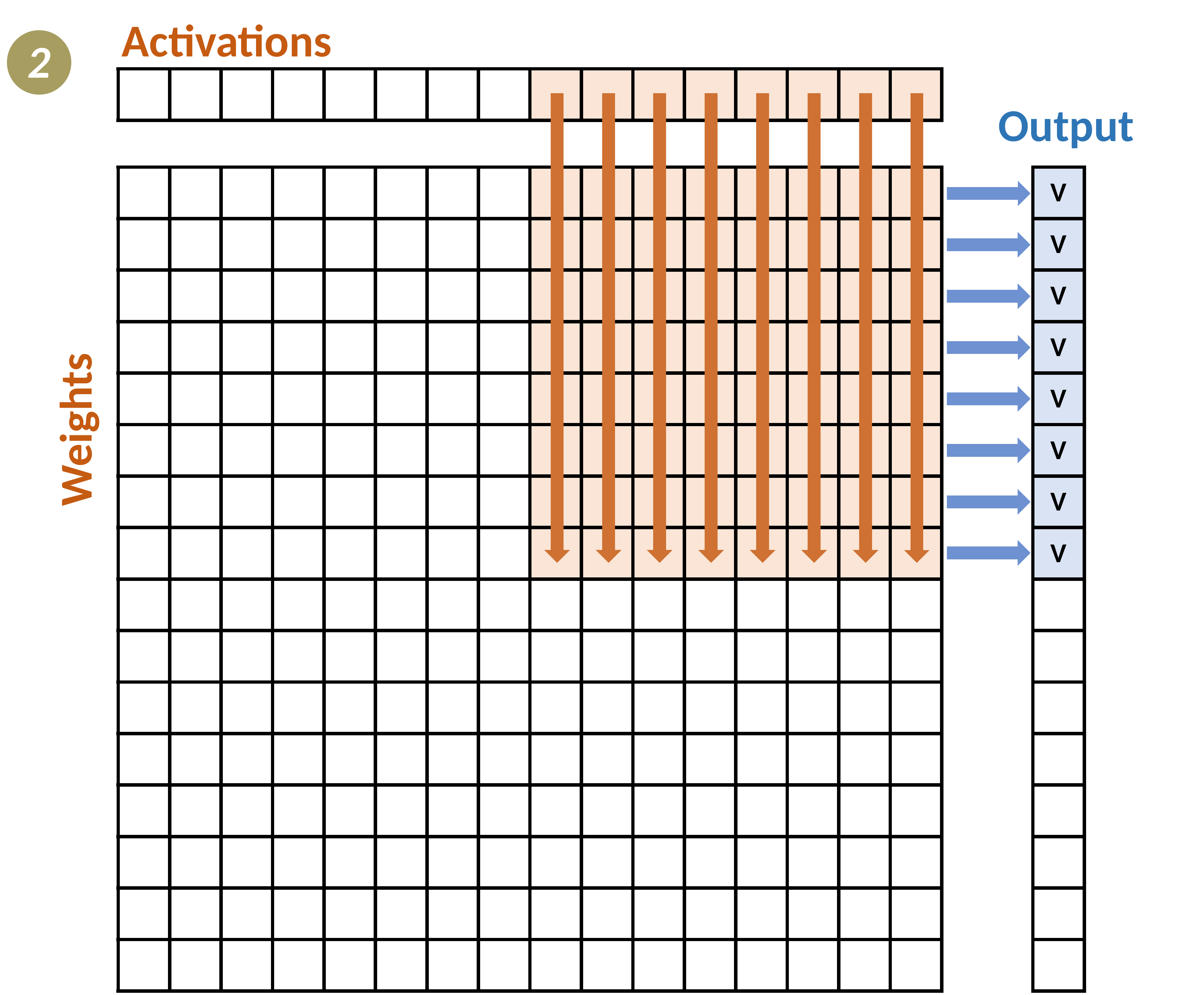}
        \label{fig:sub2}
    \end{subfigure}
    \\
    \begin{subfigure}[b]{0.23\textwidth}
        \centering
        \includegraphics[width=\textwidth]{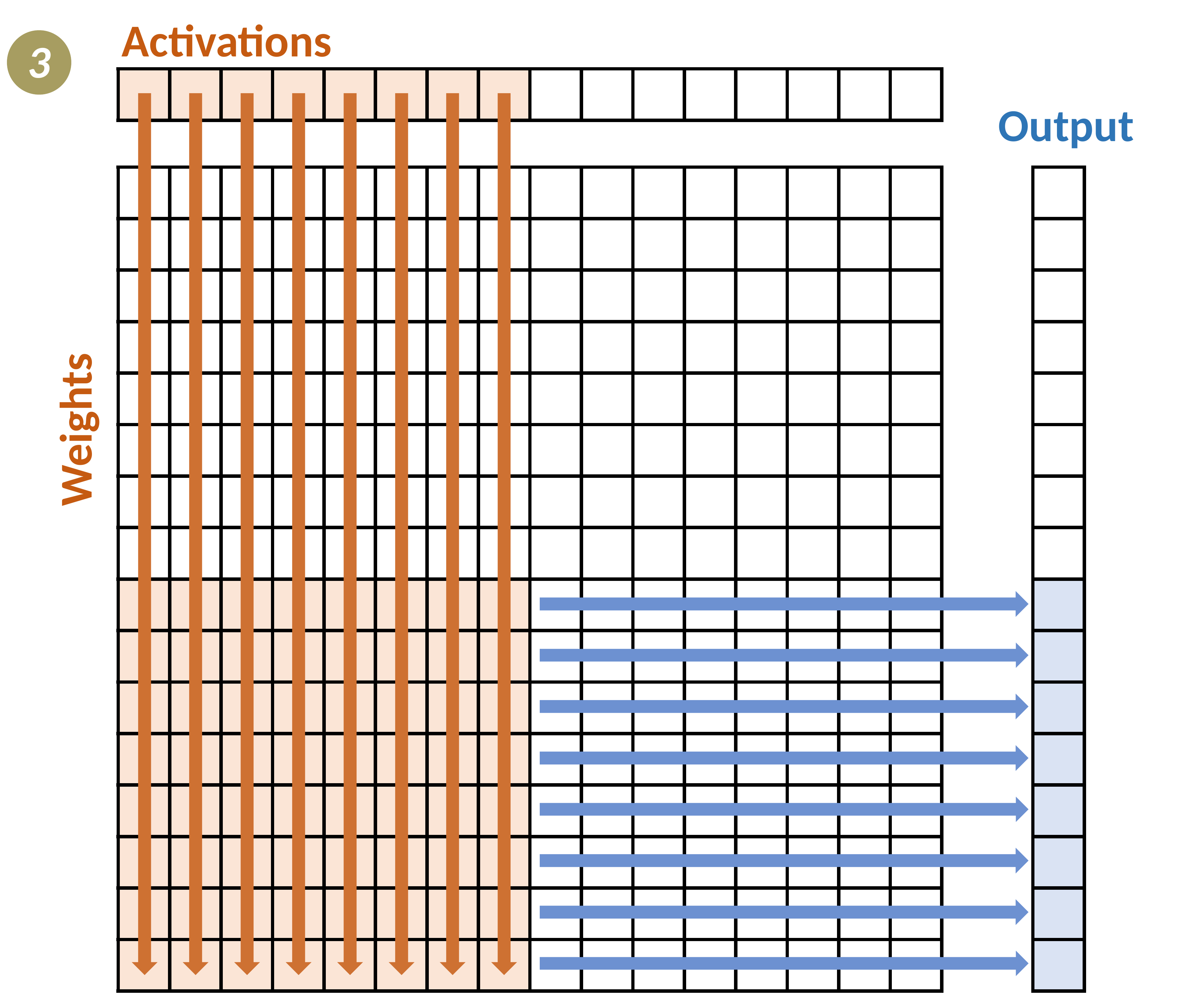}
        \label{fig:sub3}
    \end{subfigure}
    \begin{subfigure}[b]{0.23\textwidth}
        \centering
        \includegraphics[width=\textwidth]{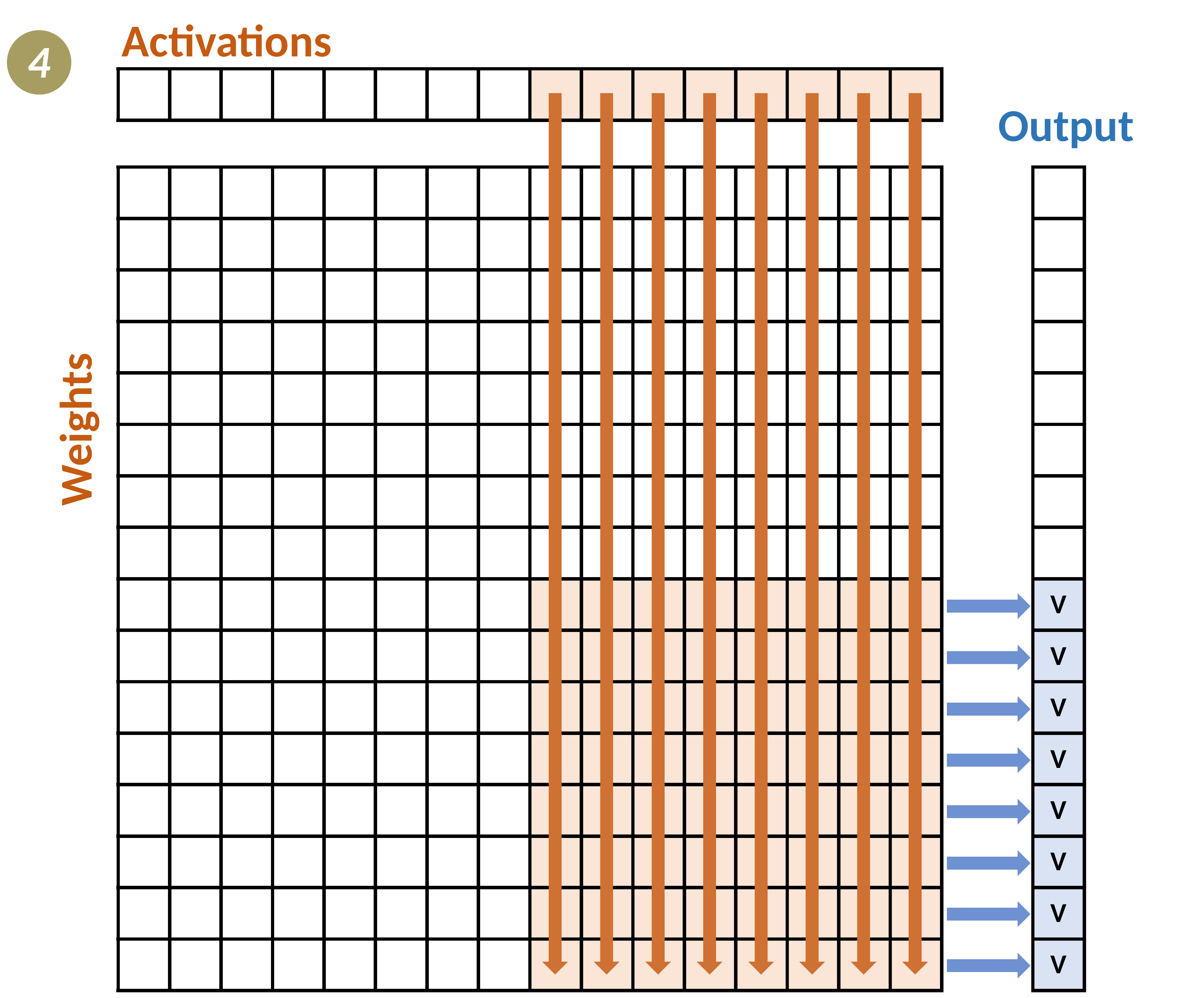}
        \label{fig:sub4}
    \end{subfigure}
    \caption{Output-stationary dataflow in TINT-Core to minimize bandwidth.}
    \label{fig:dataflow}
\end{figure}

The TINT-Core adopts an \textbf{output-stationary (OS)} dataflow (Fig.~\ref{fig:dataflow}) to minimize data movement. \revised{The activation input is broadcast to PEs from the activation buffer. Each ternary weight is connected to each PE.} In this scheme, partial sums are accumulated locally within the DFFs until the computation for a tile is complete. This dataflow effectively keeps partial sums on-chip, reducing the bandwidth consumption associated with reading and writing intermediate results.

\subsection{BoothFlex-Core Design}
\label{subsec:boothflex_core}

To handle the high-precision INT8$\times$INT8 matrix multiplications required by the attention mechanism (Query-Key and Score-Value computations) without dedicating area-inefficient INT8 multipliers, we propose the BoothFlex-Core. This unified engine executes both INT8$\times$INT8 and Ternary$\times$INT8 operations on a shared datapath.

\subsubsection{Architecture}
\begin{figure}[t]
    \centering
    \includegraphics[width=0.92\linewidth]{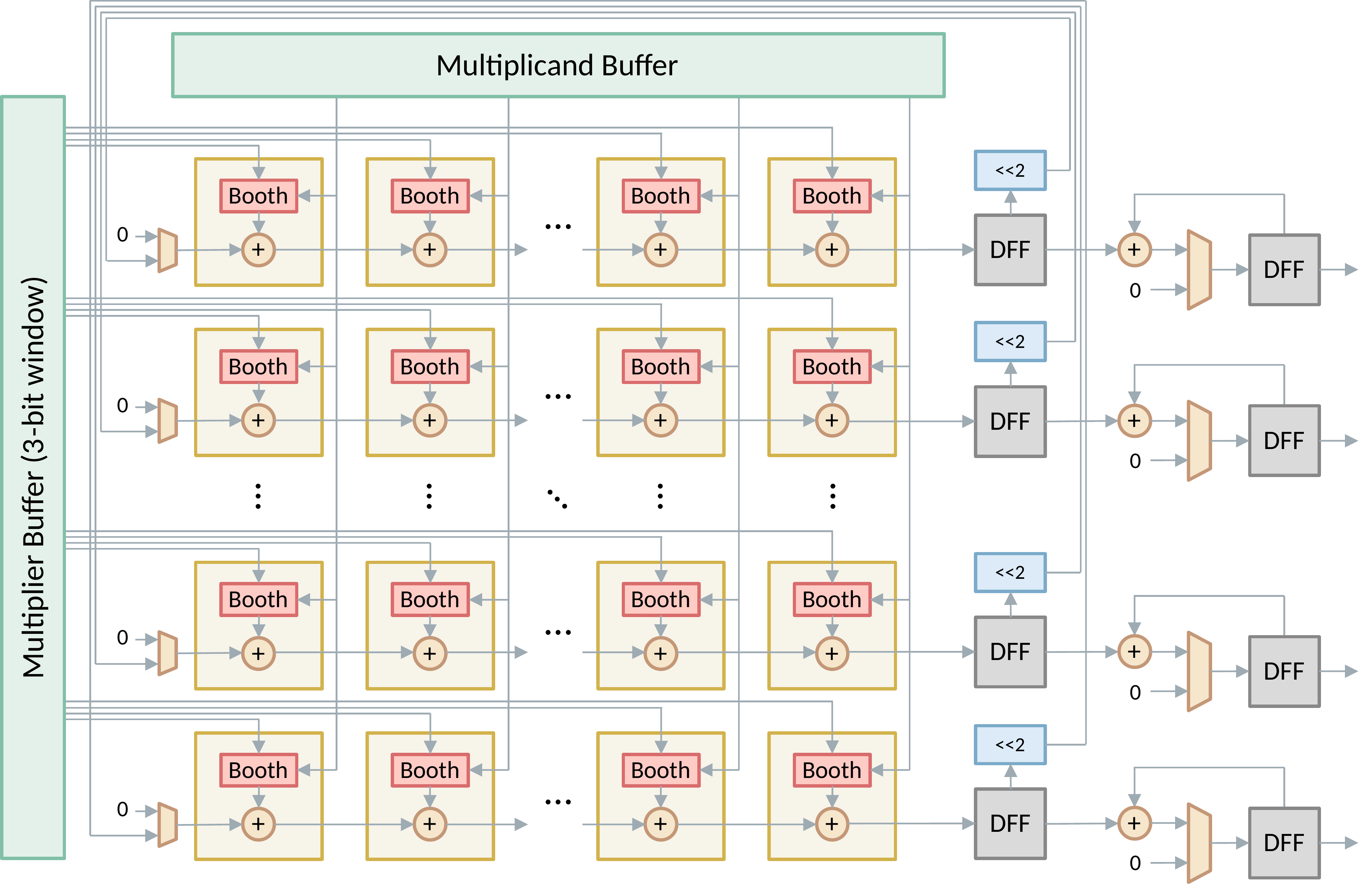}
    \caption{Architecture of the BoothFlex-Core. The Radix-4 Booth multiplier array supports both multi-cycle INT8 accumulation and single-cycle ternary projection.}
    \label{fig:boothflex_architecture}
\end{figure}

The BoothFlex-Core is an 8$\times$8 PE array based on Radix-4 Booth multipliers (Fig.~\ref{fig:boothflex_architecture}). The key architectural insight is that the arithmetic logic required for Booth encoding—which processes operands via scanning windows—can be repurposed for ternary arithmetic. By sharing the Booth encoders across both modes, the core avoids duplicating hardware arrays. To streamline system integration and control logic, BoothFlex-Core adopts the same output-stationary (OS) dataflow as the TINT-Core.

\subsubsection{Radix-4 Booth Encoding and Padding}
\label{sec: Radix-4 Booth Encoding and Padding}
Radix-4 Booth encoding reduces partial products by scanning overlapping 3-bit windows. For multiplier $X$ and multiplicand $Y$:
\begin{equation}
    Y \times X = \sum_{i=0}^{N/2-1} Y \cdot \underbrace{(-2x_{2i+1} + x_{2i} + x_{2i-1})}_{\text{Booth Factor}} \cdot 2^{2i}
\end{equation}
To support ternary weights, we employ a zero-padding mechanism where the 2-bit ternary code is mapped to a valid 3-bit Booth window by appending a logic $0$ to the LSB (Table~\ref{tab:booth_tables}). This forces the Booth logic to perform simple addition, subtraction, or zeroing, mimicking TINT-Core behavior.

\begin{table}[h]
    \centering
    \caption{Comparison of Standard Booth Encoding vs. Proposed Ternary Padding.}
    \label{tab:booth_tables}
    \small
    \renewcommand{\arraystretch}{1.1}
    \begin{subtable}{0.48\linewidth}
        \centering
        \caption{Standard Radix-4 Booth}
        \begin{tabular}{|ccc|c|}
            \hline
            \rowcolor{gray!20} $x_{2i+1}$ & $x_{2i}$ & $x_{2i-1}$ & \textbf{Op} \\ \hline
            0 & 0 & 0 & $0$ \\ 
            0 & 0 & 1 & $+Y$ \\ 
            0 & 1 & 0 & $+Y$ \\ 
            0 & 1 & 1 & $+2Y$ \\ 
            1 & 0 & 0 & $-2Y$ \\ 
            1 & 0 & 1 & $-Y$ \\ 
            1 & 1 & 0 & $-Y$ \\ 
            1 & 1 & 1 & $0$ \\ \hline
        \end{tabular}
    \end{subtable}
    \hfill
    \begin{subtable}{0.48\linewidth}
        \centering
        \caption{Ternary to Booth Inputs}
        \begin{tabular}{|c|c|c|}
            \hline
            \rowcolor{gray!20} \textbf{Weight} & \textbf{Stored} & \textbf{Padded} \\ \hline
            $+1$ & 2'b01 & 3'b01\textcolor{red}{0} \\ \hline
            $0$  & 2'b00 & 3'b00\textcolor{red}{0} \\ \hline
            $-1$ & 2'b11 & 3'b11\textcolor{red}{0} \\ \hline
            \multicolumn{3}{c}{} \\ 
            \multicolumn{3}{c}{} \\ 
            \multicolumn{3}{c}{} \\ 
            \multicolumn{3}{c}{} \\ 
            \multicolumn{3}{c}{} \\ 
        \end{tabular}
    \end{subtable}
\end{table}

\subsubsection{Bit-Serial Accumulation}
The core performs bit-serial accumulation governed by:
\begin{equation}
    \text{PartialSum}_i = \text{PartialSum}_{i-1} \cdot 2^2 + \sum_{j=0}^{7} PP_{i,j}
\end{equation}
The core dynamically adjusts iterations based on precision mode:
\begin{itemize}
    \item \textbf{Ternary$\times$INT8 Mode ($N=1$):} Completes in one cycle, matching TINT-Core throughput.
    \item \textbf{INT8$\times$INT8 Mode ($N=5$):} Performs 5 iterations ($N=\lceil \frac{8+2}{2} \rceil$) for high-precision attention.
\end{itemize}
This flexibility allows BoothFlex-Core to serve as a high-precision engine during attention phases and switch to a high-throughput engine during FFN phases, maximizing hardware utilization.

%% file: chapters/3SystemIntegrationandScheduling.tex
\section{System Integration and Scheduling}
\label{sec:system_integration}

Realizing end-to-end efficiency for BitNet b1.58 on resource-constrained edge devices requires a rigorous system-level strategy to address two critical bottlenecks: the memory bandwidth saturation caused by massive Key-Value (KV) cache traffic, and pipeline stalls induced by data-dependent operations (e.g., Softmax, RMSNorm, quantization). To overcome these challenges, we introduce three key innovations. First, we propose a \textbf{Leading One Prediction (LOP)} mechanism with a unified predictor that minimizes redundant memory accesses by identifying critical tokens prior to KV cache fetching. Second, we implement a \textbf{Head-Level Pipelining} strategy to maximize the overlap between TINT-Cores (linear projections) and BoothFlex-Core (attention). Finally, we develop a \textbf{Dependency-Aware Scheduling} methodology comprising \textit{Two-stage Nonlinear Operations} and \textit{Q-Friendly Two-Level Scheduling} to resolve pipeline stalls.

\subsection{Leading One Prediction (LOP) for Sparse Attention}
\label{subsec:lop}

\revised{The autoregressive decoding stage imposes a substantial memory-bandwidth burden due to the repeated fetching of the KV cache. Motivated by the observation that attention scores are often sparse~\cite{fact2023, sofa2024}, we propose a hardware-efficient \textbf{Leading One Prediction (LOP)} mechanism to reduce unnecessary off-chip memory accesses. Unlike FACT~\cite{fact2023}, which prunes projections in the prefill stage, our design adopts a \textbf{Unified Predictor} that supports both parallel prefill and sequential decode execution under a shared prediction framework. In our implementation, the same LOP formulation is applied uniformly across all attention heads and layers, avoiding the need for layer-specific predictor designs or runtime reconfiguration. Furthermore, given the low computational cost of ternary projections in BitNet b1.58, our design focuses exclusively on pruning the expensive KV-cache memory accesses rather than the projections themselves, thereby improving memory efficiency with minimal disruption to the original computation flow.}

\subsubsection{Unified Prediction Algorithm}
The core concept is to estimate the dot-product similarity between a Query ($q$) and Key ($k$) using the position of their leading one bits ($LO(x) = \lfloor \log_2 |x| \rfloor$). The surrogate score $S(q, k)$ is calculated as:
\begin{equation}
    \label{eq:simplified_score}
    S(q, k) = \sum_{i=1}^{d} \text{sgn}(q_i)\text{sgn}(k_i) \cdot 2^{LO(q_i) + LO(k_i)}
\end{equation}
To facilitate hardware efficiency, INT8 values are compressed into a compact 4-bit representation consisting of a single Sign bit and a 3-bit Leading One (LO) position. This compression allows replacing complex multipliers with simple shift-and-add operations and significantly reduces the bandwidth required for prediction. By multiplying these low-precision $Q_{LO}$ and $K_{LO}$ features, the system generates surrogate scores to identify a sparse set of top candidates, fetching full-precision $K$ and $V$ vectors only for these indices.

\subsubsection{Hardware Implementation}
\begin{figure}[t]
    \centering
    \includegraphics[width=0.98\linewidth]{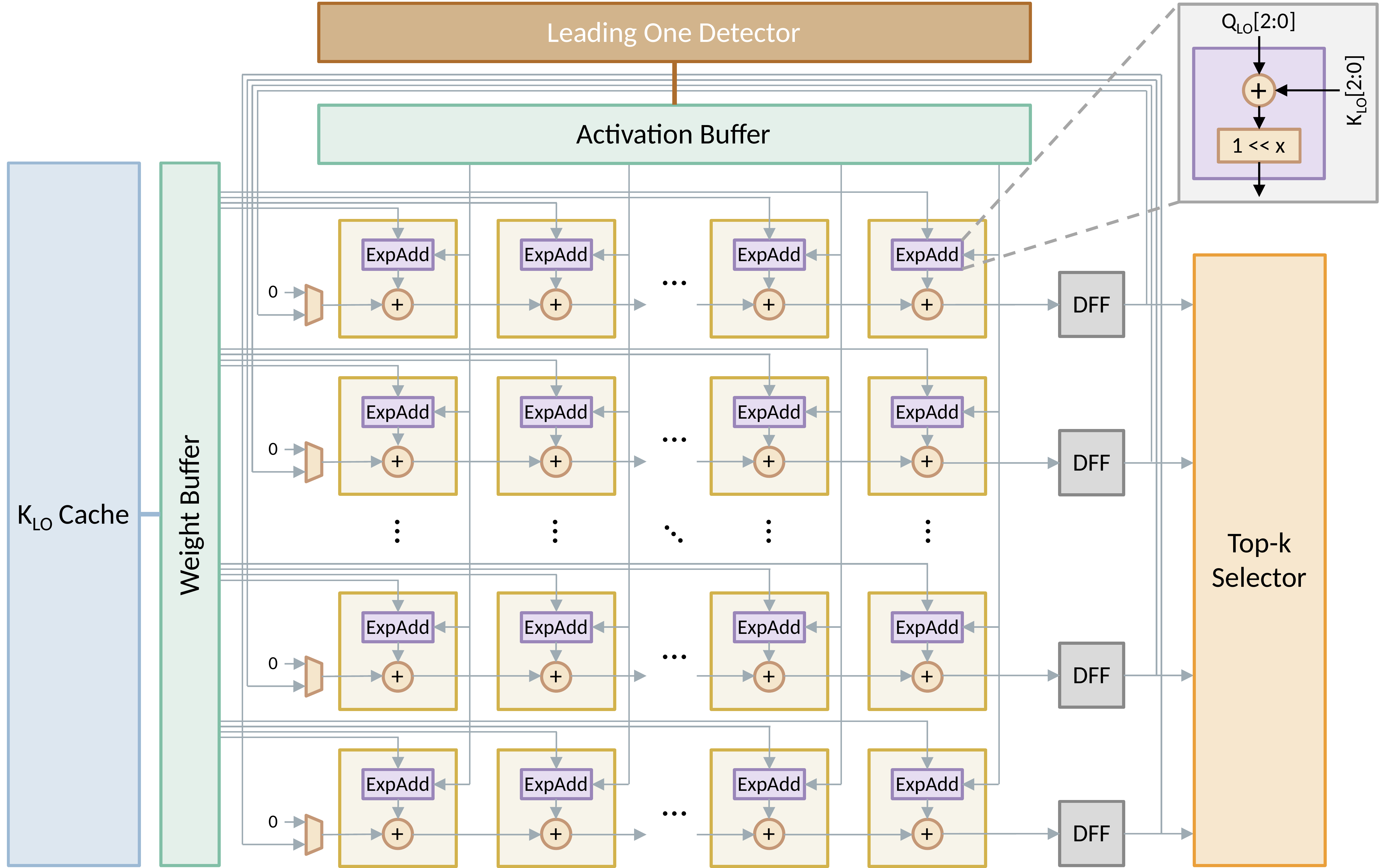}
    \caption{Microarchitecture of LOP-Core featuring ExpAdd arrays and a Top-K Selector.}
    \label{fig:lop_architecture}
\end{figure}

To implement this algorithm efficiently, we design a dedicated LOP-Core. As illustrated in Fig.~\ref{fig:lop_architecture}, the core features an array of \textbf{ExpAdd} units (composed of shifters and adders) that compute surrogate scores in parallel. The resulting scores are streamed to a \textbf{Comparison-free Top-K Selector} based on bitwise logic~\cite{saha2023kdegree}, which filters the indices of the highest scores without high-latency sorting comparisons.

This mechanism reduces KV cache traffic by approximately $1 - 32/M$ (where $M$ is the sequence length) by fetching only the top-32 relevant tokens. We evaluate the impact of LOP on model convergence using the \textbf{WikiText-2} dataset. Experimental results demonstrate that this aggressive pruning strategy yields a perplexity of $10.15135$. This performance is virtually identical to the \textbf{dense baseline} (i.e., execution without LOP), which achieves $10.15090$, confirming that the proposed mechanism maintains model quality with negligible degradation.

\subsection{Head-Level Pipelining}
\label{subsec:head_level_pipelining}

Standard layer-wise execution enforces a strict dependency: linear projections for \textit{all} heads must complete before attention computation begins. This imposes two major penalties: (1) \textbf{Computation Bubbles}, as the attention engine sits idle during the projection phase; and (2) \textbf{Memory Overheads}, as the system must buffer QKV tensors for the entire layer ($H \times d_k \times SeqLen$), often exceeding on-chip SRAM capacity.

\begin{figure}[t]
    \centering
    \includegraphics[width=1.0\linewidth]{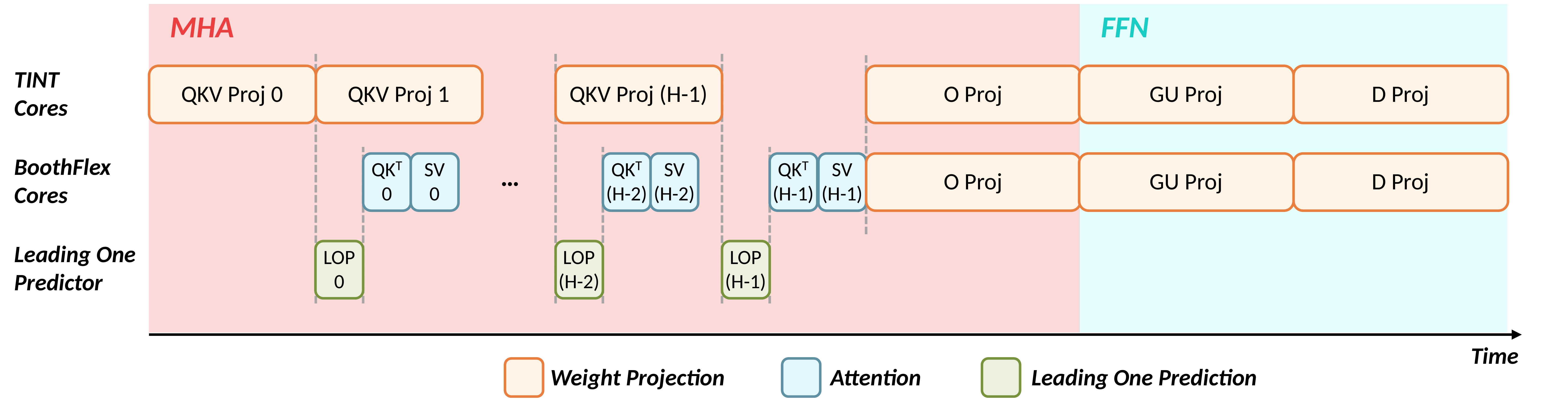}
    \caption{Timeline of the Head-Level Pipelining strategy. By overlapping the weight projections of Head $h$ (on TINT-Cores) with the attention computation of Head $h-1$ (on BoothFlex-Core), the design effectively hides the attention latency.}
    \label{fig:head_pipeline}
\end{figure}

To resolve these inefficiencies, drawing inspiration from \textbf{Energon}~\cite{energon2022} and \textbf{AttAcc}~\cite{attacc2024}, we propose an adapted \textbf{Head-Level Pipelining} strategy (Fig.~\ref{fig:head_pipeline}). Unlike standard layer-wise execution, this approach breaks the synchronization barrier and orchestrates execution at the finer granularity of a single attention head, maximizing the overlap between heterogeneous cores.

\subsubsection{Pipeline Strategy}
We treat the TINT-Cores as producers and the BoothFlex-Core as a consumer, overlapping their execution:
\begin{itemize}
    \item \textbf{Attention Phase:} While TINT-Cores compute projections ($Q_h, K_h, V_h$) for the current head $h$, the BoothFlex-Core simultaneously executes the attention mechanism ($QK^T$ and $SV$) for the previously completed head $h-1$.
    \item \textbf{FFN Phase:} Upon completing attention, the BoothFlex-Core switches to ternary mode to assist TINT-Cores with the massive $W_O$ and FFN projections, ensuring high utilization.
\end{itemize}

\subsubsection{Memory Optimization}
Unlike non-pipelined schedules that buffer tensors for all $H$ heads, our streaming strategy consumes Q/K/V vectors immediately. This reduces the lifecycle of intermediate tensors to a single head's processing time. Consequently, the on-chip storage requirement is drastically reduced to buffering only two heads (one being produced, one being consumed), allowing the design to fit within a compact 73.14 KB SRAM.

\subsection{Dependency-Aware Scheduling}
\label{subsec:dependency_aware}

Data-dependent operations, such as Softmax, RMSNorm, and quantization, inherently require global reductions (e.g., global sum or maximum) across the entire vector. In a naive implementation, this creates a dependency barrier where the pipeline must stall until the full vector is computed, leading to significant hardware underutilization. To resolve this, we propose a scheduling framework comprising \textit{Two-stage Nonlinear Operations} and \textit{Q-Friendly Two-Level Scheduling}.

\subsubsection{Two-Stage Nonlinear Operations with Deferred Dependency}
\begin{figure}[t]
    \centering
    \begin{subfigure}[b]{0.98\linewidth}
        \centering
        \includegraphics[width=\textwidth]{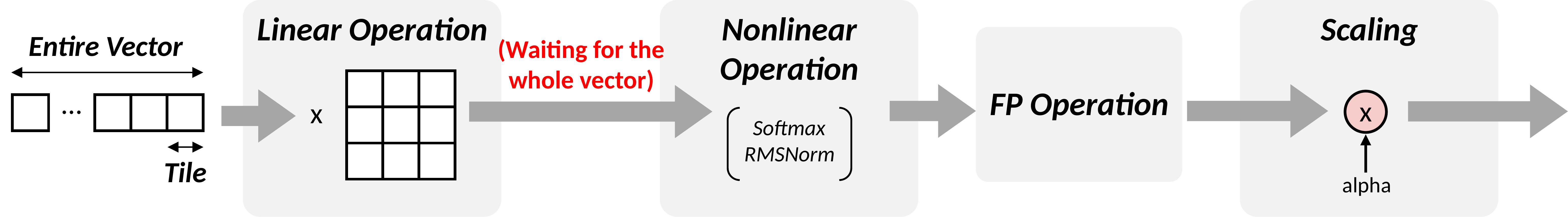}
        \caption{Dependency barrier in conventional scheduling.}
        \label{fig:nonlinear_dependency}
    \end{subfigure}
    \vspace{0.5em}
    \begin{subfigure}[b]{0.98\linewidth}
        \centering
        \includegraphics[width=\textwidth]{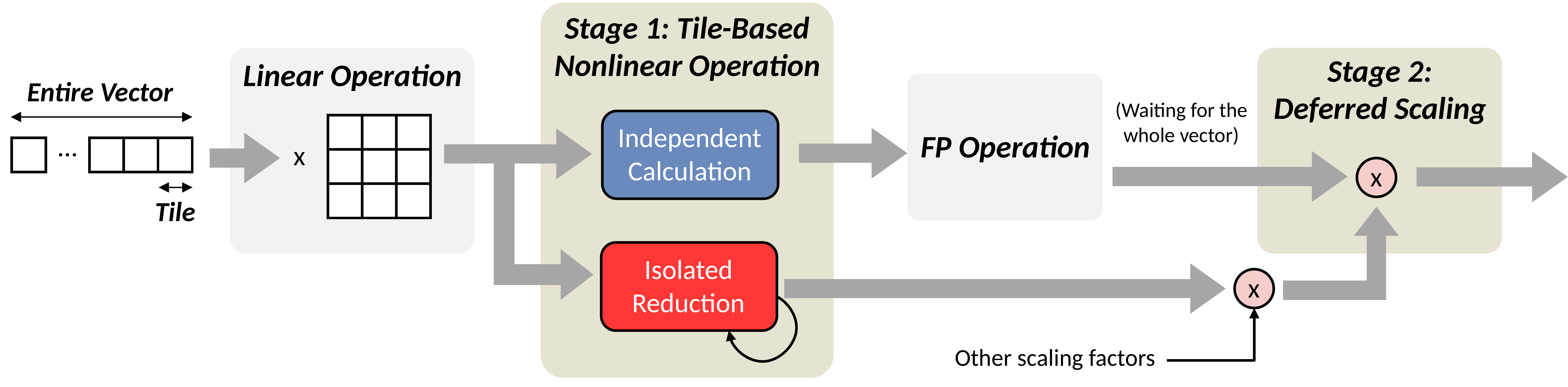}
        \caption{Proposed Two-stage strategy with deferred scaling.}
        \label{fig:two_stage_nonlinear}
    \end{subfigure}
    \caption{Comparison of dependency handling strategies. The proposed design hides the latency of global reductions by deferring the final scaling.}
    \label{fig:nonlinear_scheduling}
\end{figure}

We resolve the blocking dependency in Softmax and RMSNorm by decomposing the execution into two stages, as illustrated in Fig.~\ref{fig:nonlinear_scheduling}.

\begin{itemize}
    \item \textbf{Stage 1 (Tile-Based):} The hardware processes incoming tiles immediately. It performs independent element-wise calculations (e.g., $x_i \cdot w_i$ for RMSNorm or $e^{x_i-M}$ for Softmax) and accumulates local partial sums in a streaming fashion, without waiting for the full vector.
    \item \textbf{Stage 2 (Deferred Scaling):} The global reduction (sum or RMS) is computed only after all tiles are processed. The final division is mathematically deferred and fused with the subsequent quantization scaling factor.
\end{itemize}

For Softmax specifically, waiting for the dynamic global maximum ($\max(x)$) creates a ``double dependency'' that poses a critical challenge: it prevents even the independent calculations in Stage 1 from starting (as $e^{x_i - \max(x)}$ cannot be computed without the max value). To bridge this gap and enable our Two-Stage strategy, we adopt the \textbf{Unified Max} strategy that is inspired by~\cite{flashdecoding++2024}. Statistical analysis of BitNet b1.58 shows that attention scores are concentrated in a small range; thus, we employ a static upper bound $M_{\text{unified}} = 16$. This conservative constant ensures numerical stability without dynamic scanning. Experimental results on the \textbf{WikiText-2} dataset verify that this approximation yields a perplexity of $10.15135$. This is virtually identical to the exact baseline ($10.15090$), confirming that using a static unified max value validates the approach without compromising model accuracy.

\subsubsection{Q-Friendly Two-Level Scheduling}
\begin{figure}[t]
    \centering
    \begin{subfigure}[b]{0.98\linewidth}
        \centering
        \includegraphics[width=\textwidth]{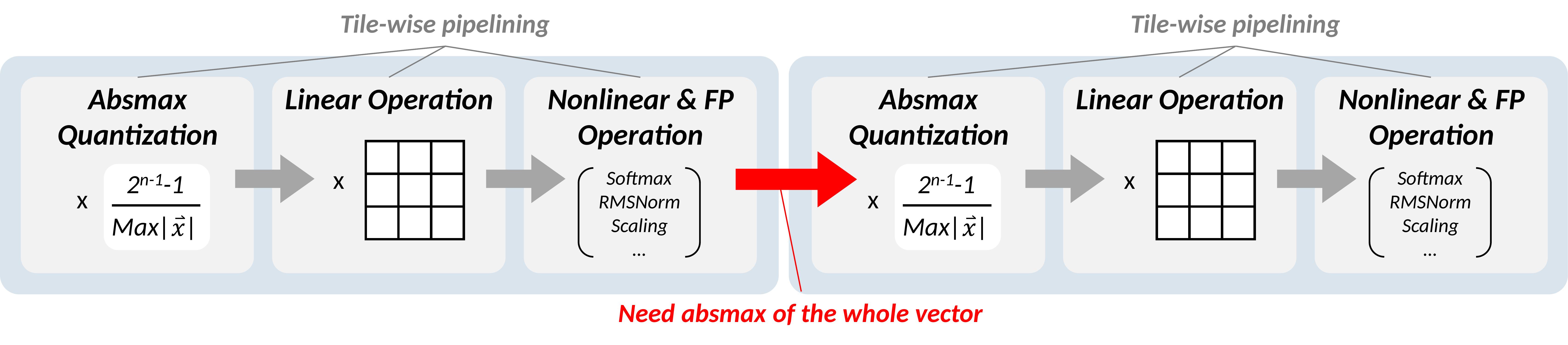}
        \caption{The Quantization Barrier requiring global maximum.}
        \label{fig:quantization_barrier}
    \end{subfigure}
    \vspace{1em} %
    \begin{subfigure}[b]{0.98\linewidth}
        \centering
        \includegraphics[width=\textwidth]{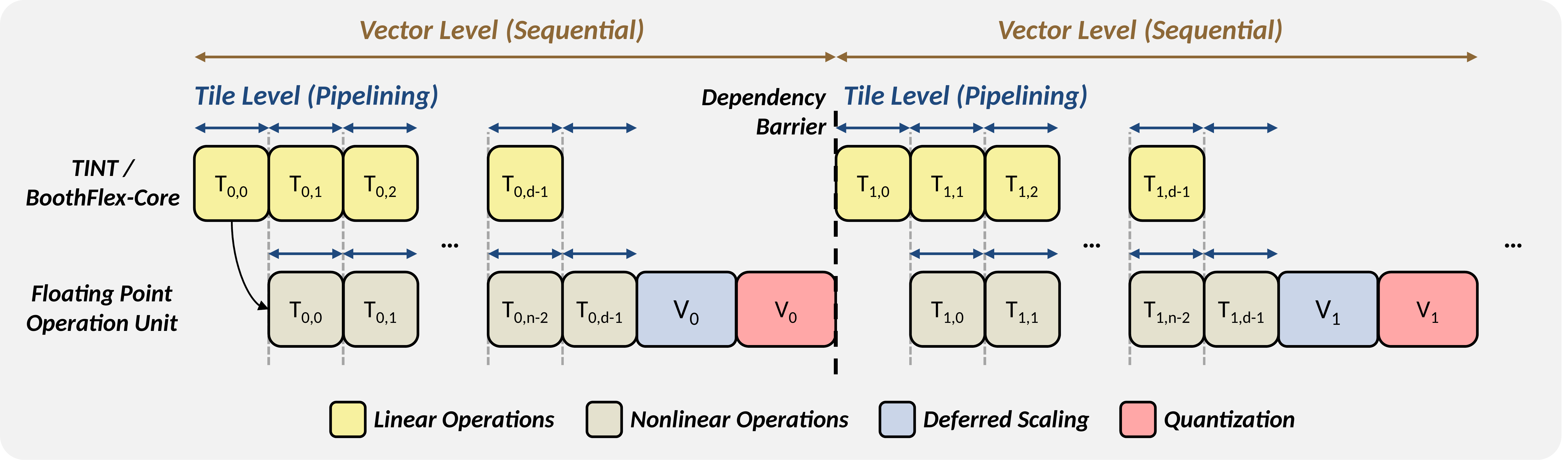}
        \caption{Q-Friendly Two-Level Scheduling hierarchy.}
        \label{fig:q_friendly_scheduling}
    \end{subfigure}
    \caption{Handling the quantization dependency. (a) The hard dependency breaks the pipeline. (b) Hierarchical scheduling maintains fine-grained pipelining within vectors while managing synchronization at the coarse-grained vector level.}
    \label{fig:quant_scheduling}
\end{figure}

The activation quantization in BitNet b1.58 requires the global absolute maximum ($\max|\vec{x}|$) to determine the scaling factor. As shown in Fig.~\ref{fig:quant_scheduling}(a), this creates a hard synchronization point that inevitably breaks the pipeline between layers.

Rather than eliminating this dependency, we propose a hierarchical scheduling strategy (Fig.~\ref{fig:quant_scheduling}(b)):
\begin{enumerate}
    \item \textbf{Vector Level (Sequential):} We enforce sequential execution between vectors. The quantization unit acts as a synchronization barrier, ensuring that the global statistics for vector $t$ are finalized before processing vector $t+1$.
    \item \textbf{Tile Level (Pipelined):} Within a single vector, linear and nonlinear operations execute in a continuous and stall-free pipeline. The quantization dependency is effectively isolated as a finalization step at the vector boundary, ensuring that the critical compute path remains unblocked by global statistics.
\end{enumerate}

This hierarchical approach ensures that while the transition between vectors is sequential, the majority of the computational workload (the tiles within a vector) remains highly pipelined and utilized.

%% file: chapters/4ExperimentalEvaluation.tex
\section{Experimental Evaluation}
\label{sec:experimental_evaluation}

We evaluate the performance of VitaLLM against state-of-the-art ASIC and FPGA accelerators in terms of throughput, area, and energy efficiency. Furthermore, we present physical implementation results, bandwidth analysis, and ablation studies to validate the proposed architectural innovations under edge constraints.

\subsection{Implementation Setup}
\label{subsec:implementation_setup}

The VitaLLM accelerator was implemented using a standard cell-based design flow in \textbf{TSMC 16nm} technology. Post-layout simulations confirm that the design operates at \textbf{1 GHz} with a supply voltage of \textbf{0.8 V}. The evaluation targets the BitNet b1.58 3B model to assess system-level performance across both prefill and decode stages.

\revised{
For the model-quality evaluation, the reported perplexity results were obtained using a bit-accurate software-level simulator that faithfully emulates the proposed hardware datapath, including the fixed-point arithmetic, LOP pruning, and the unified max approximation. This evaluation corresponds to end-to-end autoregressive inference under the same quantization and approximation settings as the proposed hardware design. The reported latency and throughput metrics are derived from a comprehensive analytical performance model that accounts for hardware parallelism, data movement overheads, and pipeline stalls based on the cycle-accurate execution flow of our architecture. To ensure physical accuracy, these analytical results are cross-verified with post-layout gate-level simulations.
}

\subsection{Physical Implementation and Bandwidth Analysis}
\label{subsec:physical_implementation}

\subsubsection{Layout and Area Breakdown}
\begin{figure}[t]
    \centering
    \includegraphics[width=0.75\linewidth]{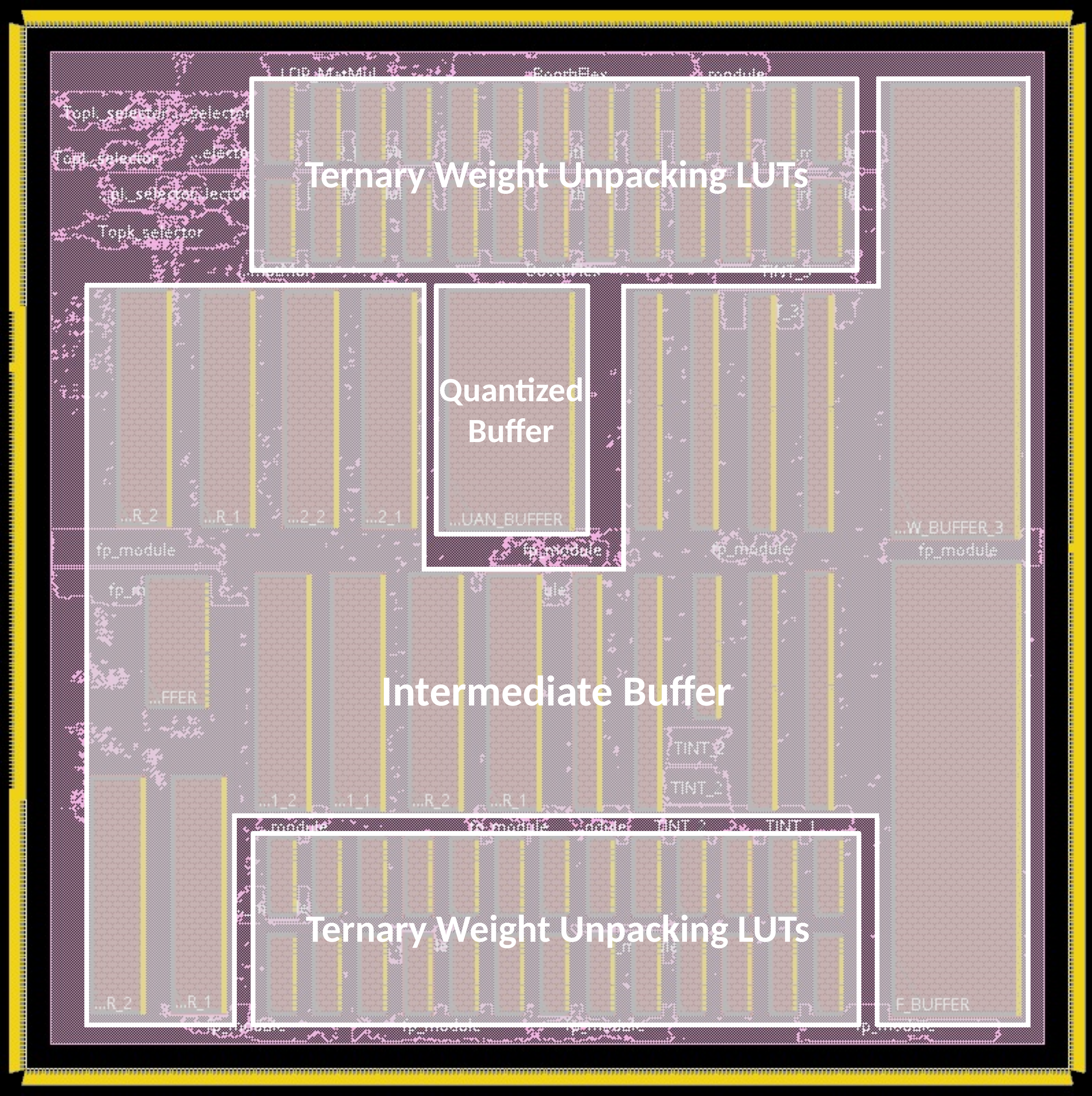}
    \caption{Physical layout of VitaLLM in TSMC 16nm ($0.223 \text{ mm}^2$).}
    \label{fig:chip_layout}
\end{figure}

The physical layout (Fig.~\ref{fig:chip_layout}) achieves a compact core area of $0.223 \text{ mm}^2$. To minimize data movement, SRAM buffers are distributed to align physically with computing units.

As shown in Fig.~\ref{fig:area_power_breakdown}(a), \textbf{on-chip SRAMs} dominate the area (69.20\%), reflecting the memory-intensive nature of LLMs despite our buffer optimizations (Table~\ref{tab:memory_breakdown}). Conversely, the computing cores (TINT-Core and BoothFlex-Core) occupy only 2.29\% combined, highlighting the density of our multiplier-free logic.

\begin{table}[h]
    \centering
    \caption{On-chip Memory Breakdown (Total: 130.5 KB).}
    \label{tab:memory_breakdown}
    \small
    \begin{tabular}{|l|c|}
        \hline
        \textbf{Component} & \textbf{Size (Bytes)} \\ \hline
        Quantized Buffer & 26,112 \\
        Ternary Weight LUTs & 16,640 \\
        Intermediate Buffer (Activation/RMS/etc.) & 87,748 \\ \hline
        \textbf{Total} & \textbf{130.5 KB} \\ \hline
    \end{tabular}
\end{table}

\subsubsection{Power Breakdown}
\begin{figure}[t]
    \centering
    \begin{subfigure}[b]{0.92\linewidth}
        \centering
        \includegraphics[width=\linewidth]{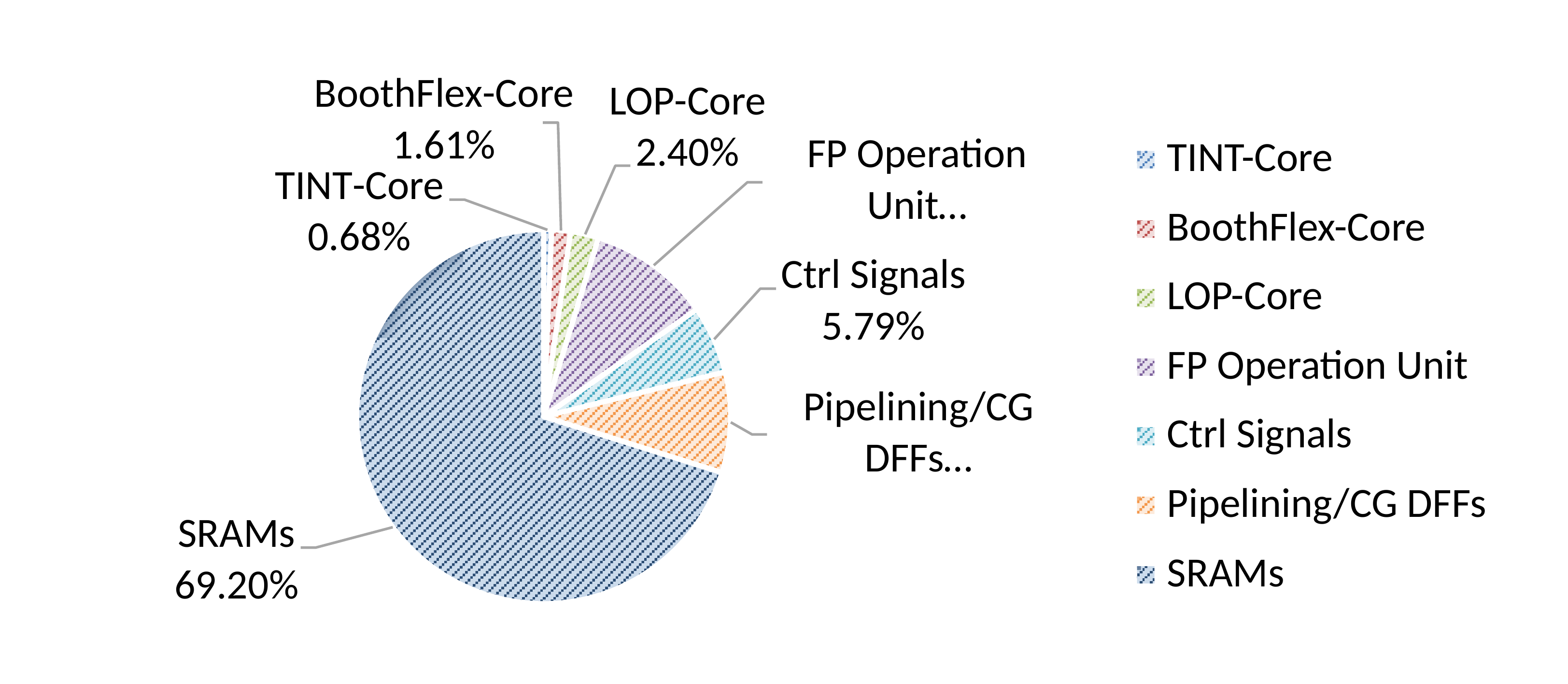}
        \caption{Area Breakdown}
    \end{subfigure}
    \vspace{1em} %
    \begin{subfigure}[b]{0.92\linewidth}
        \centering
        \includegraphics[width=\linewidth]{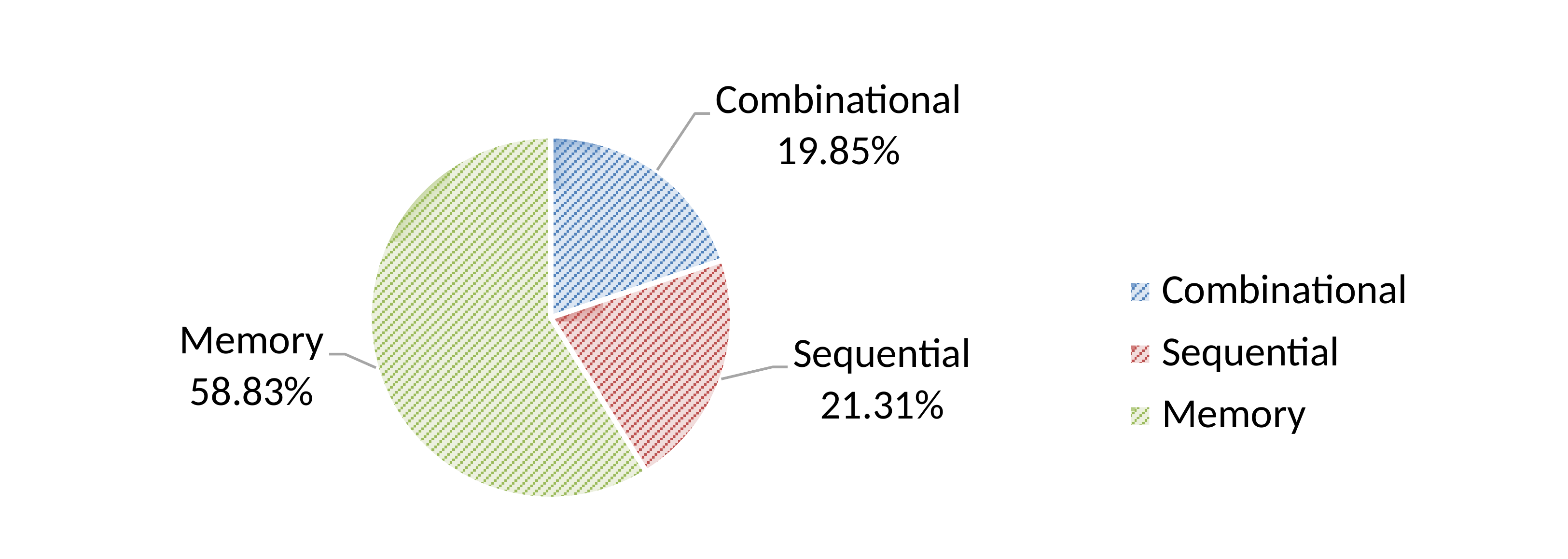}
        \caption{Power Breakdown}
    \end{subfigure}
    \caption{(a) SRAMs dominate area (69.20\%). (b) Memory operations dominate power (58.83\%).}
    \label{fig:area_power_breakdown}
\end{figure}

Fig.~\ref{fig:area_power_breakdown}(b) reveals that memory operations consume \textbf{58.83\%} of the total power. This confirms that data movement is the primary energy bottleneck, validating our choice of an output-stationary dataflow to minimize intermediate access.

\subsubsection{Bandwidth Analysis}
We target an LPDDR5T system (76.8 GB/s). As detailed in Table~\ref{tab:bandwidth_analysis}, the peak demand occurs during the Read phase of the Decode stage (\textbf{63 GB/s}), driven by non-shared ternary weights. This is well within the 76.8 GB/s limit. Notably, the minimum bandwidth required for 20 tokens/s is only \textbf{17.80 GB/s}, indicating significant adaptability to lower-end platforms.

\begin{table}[h]
    \centering
    \caption{Bandwidth Analysis (Peak vs. Limit).}
    \label{tab:bandwidth_analysis}
    \small
    \begin{tabular}{|c|c|c|}
        \hline
        \textbf{Stage} & \textbf{Peak Read} & \textbf{Peak Write} \\ \hline
        Prefill & 37 GB/s & 36 GB/s \\ 
        Decode & \textbf{63 GB/s} & 21 GB/s \\ \hline
        \textbf{Limit} & \multicolumn{2}{c|}{\textbf{76.8 GB/s (LPDDR5T)}} \\ \hline
    \end{tabular}
\end{table}

\subsection{Ablation Studies}
\label{subsec:ablation}

We quantify the contributions of the proposed techniques through ablation studies.

\subsubsection{Leading One Prediction (LOP)}
\begin{figure}[t]
    \centering
    \begin{subfigure}[b]{0.48\linewidth} 
        \centering
        \includegraphics[width=\linewidth]{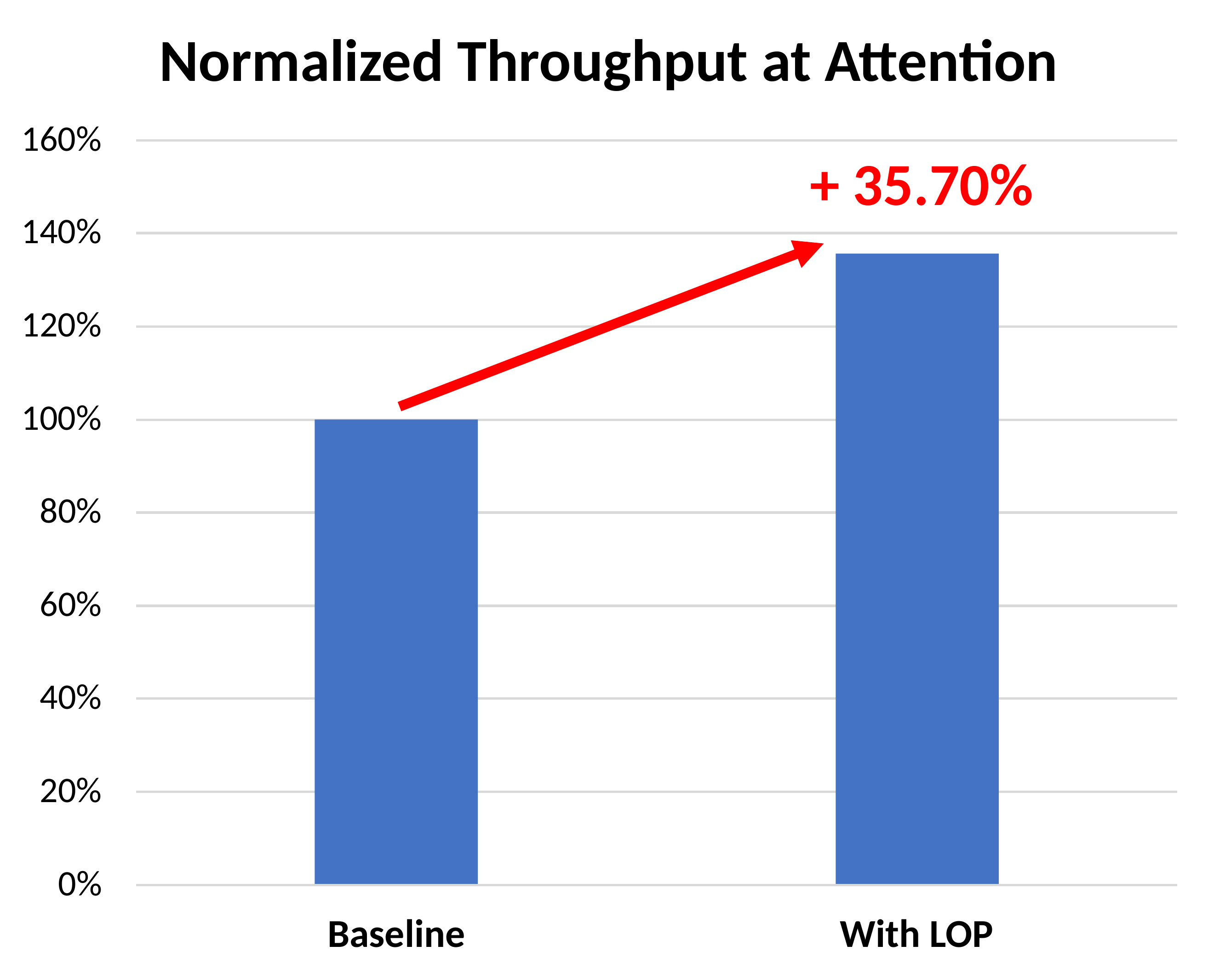}
        \caption{Throughput}
    \end{subfigure}
    \hfill
    \begin{subfigure}[b]{0.48\linewidth}
        \centering
        \includegraphics[width=\linewidth]{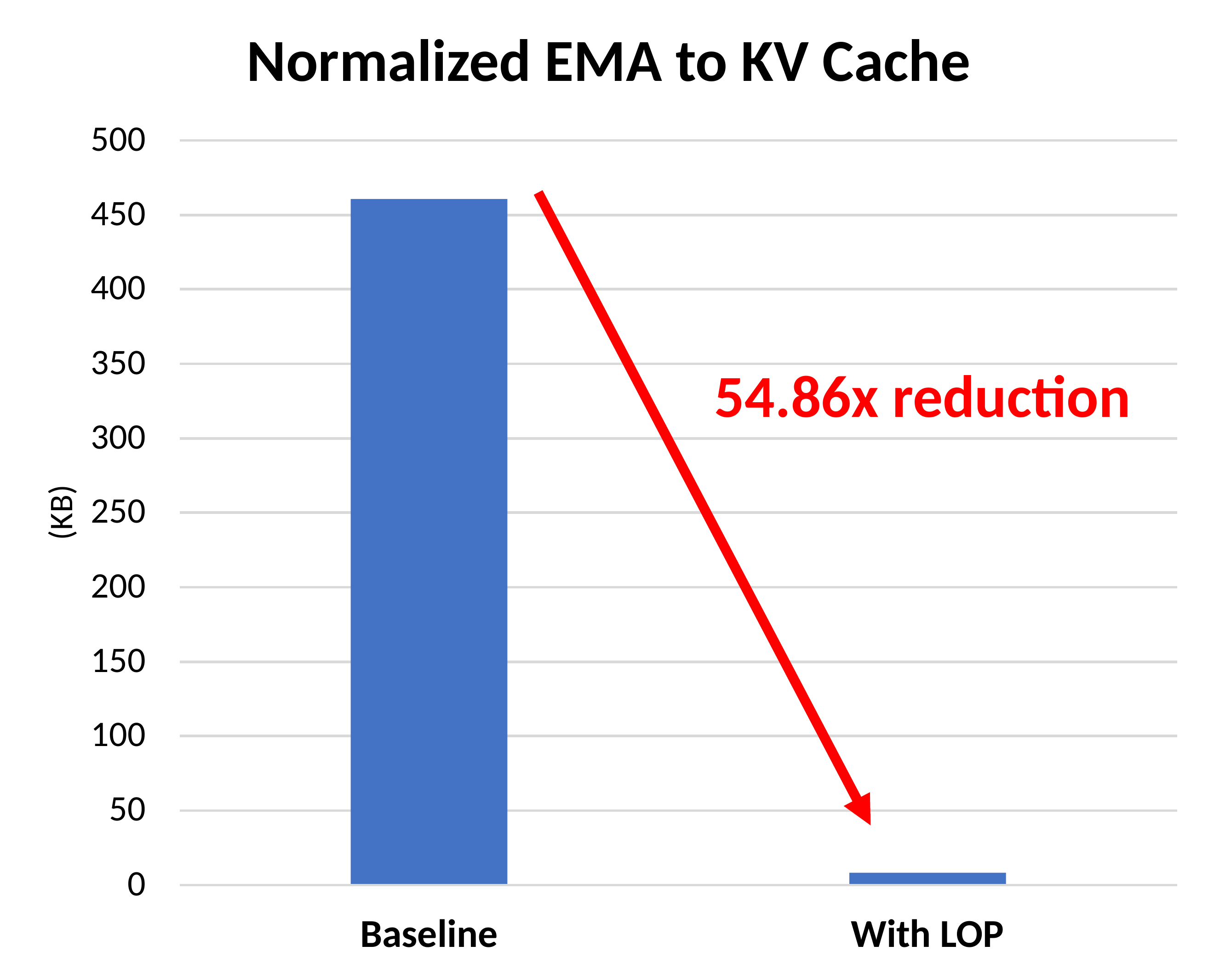}
        \caption{EMA Reduction}
    \end{subfigure}
    \caption{Impact of LOP: (a) +35.70\% throughput, (b) 54.86$\times$ less EMA.}
    \label{fig:lop_efficacy}
\end{figure}

Fig.~\ref{fig:lop_efficacy} demonstrates the efficacy of LOP in the decode stage. By pruning redundant KV cache fetches, LOP achieves a \textbf{54.86$\times$ reduction} in external memory access (EMA) and improves attention throughput by \textbf{35.70\%}.

\subsubsection{Dependency-Aware Scheduling}
\begin{figure*}[t]
    \centering
    \begin{subfigure}[htb]{0.23\linewidth}
        \centering
        \includegraphics[width=\linewidth]{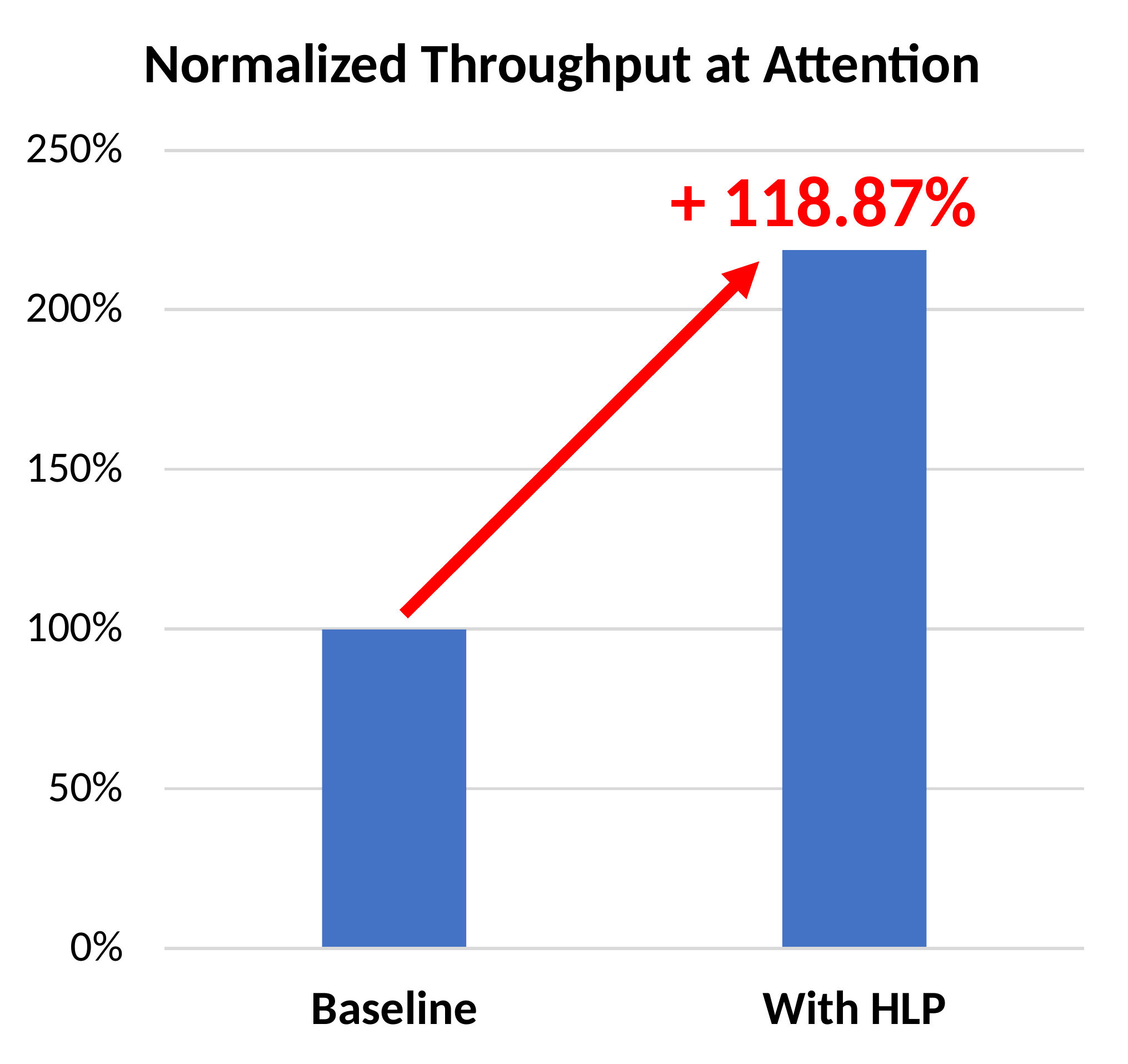}
        \caption{Attention Throughput}
        \label{fig:throughput_attention}
    \end{subfigure}
    \hspace{1em}
    \begin{subfigure}[htb]{0.23\linewidth}
        \centering
        \includegraphics[width=\linewidth]{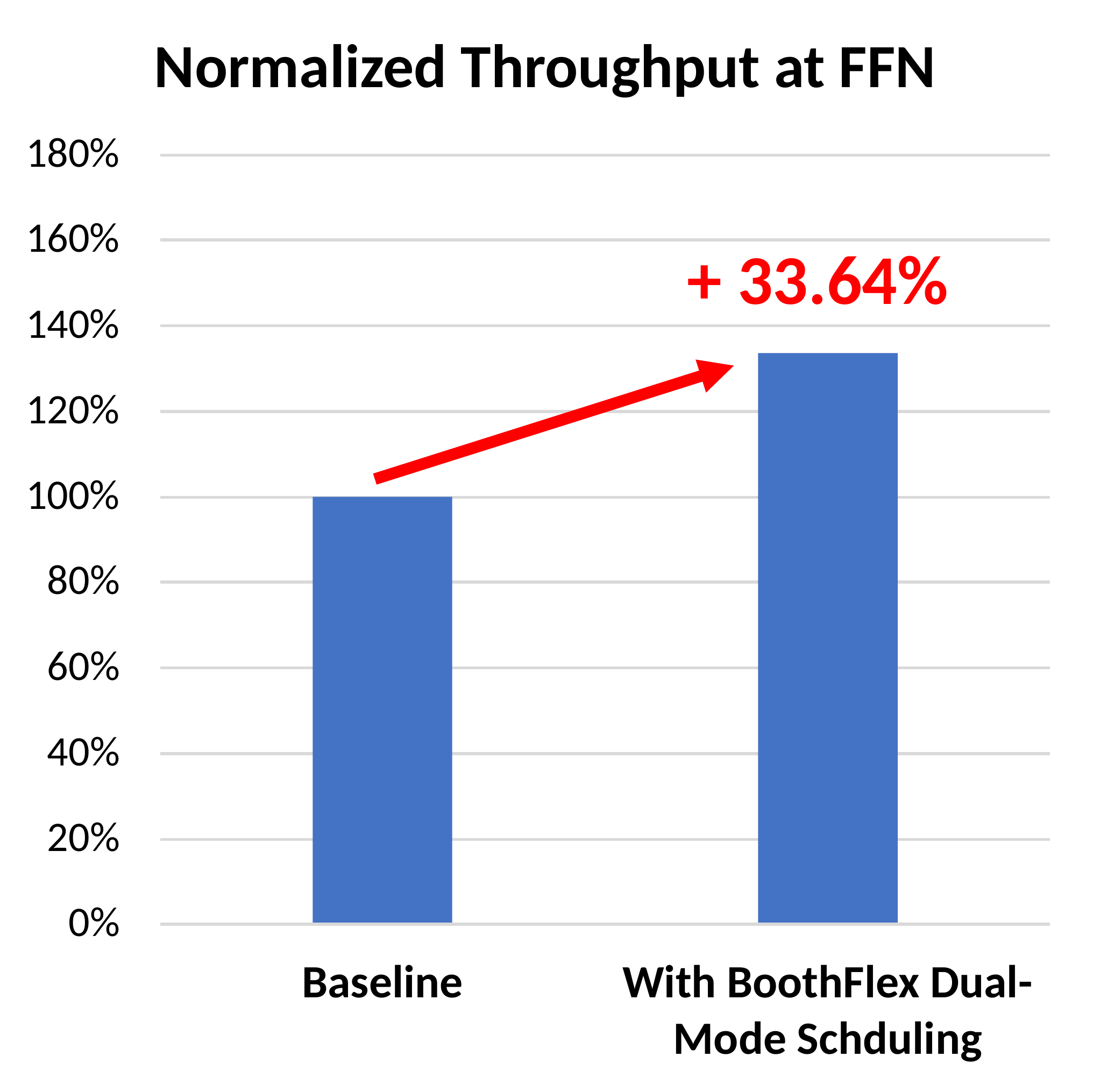}
        \caption{FFN Throughput}
        \label{fig:throughput_ffn}
    \end{subfigure}
    \hspace{1em}
    \begin{subfigure}[htb]{0.23\linewidth}
        \centering
        \includegraphics[width=\linewidth]{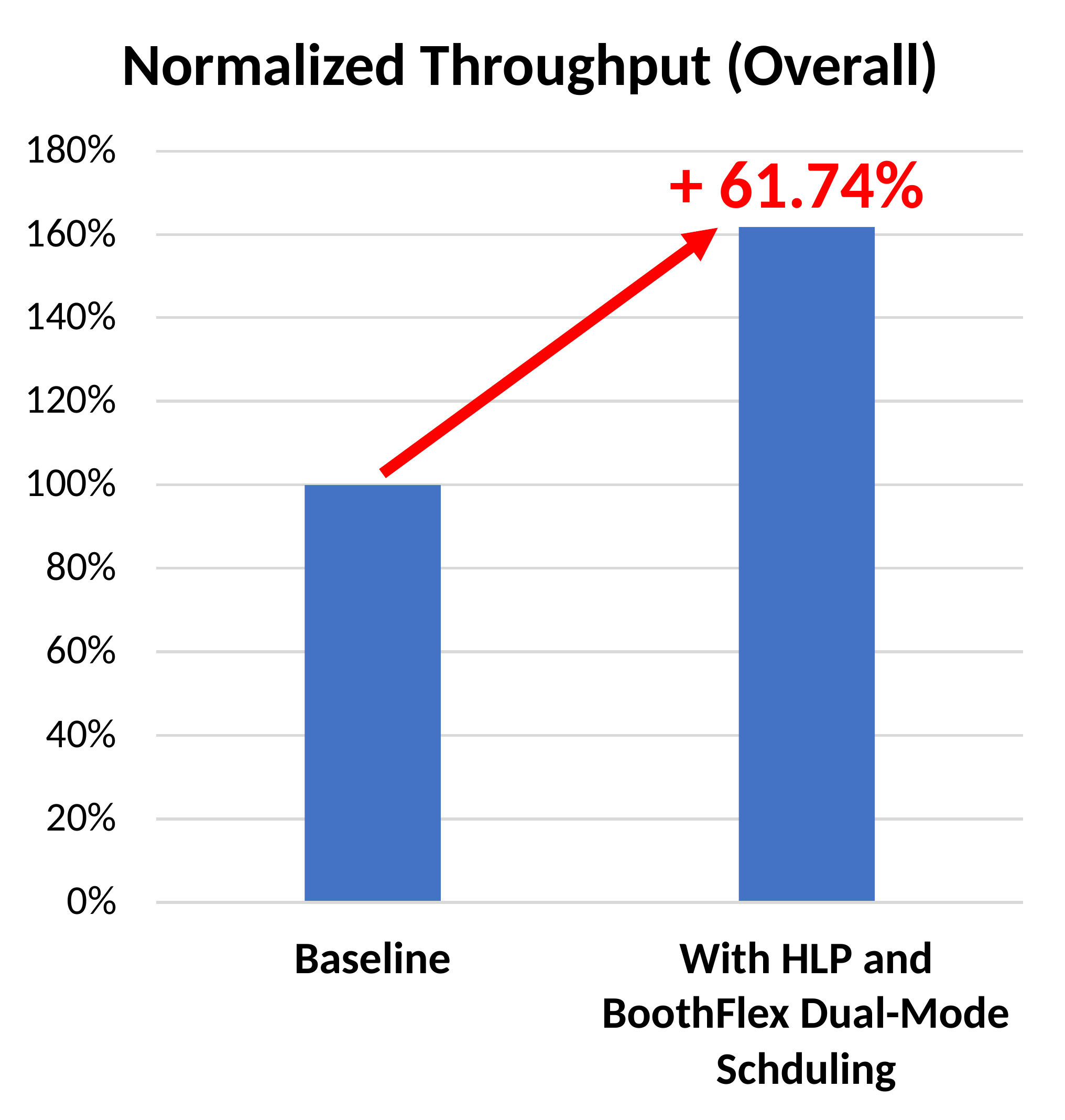}
        \caption{Overall Throughput}
        \label{fig:throughput_overall}
    \end{subfigure}
    
    \caption{Normalized throughput improvements. (a) Head-Level Pipelining (HLP) improves Attention throughput by 118.87\%. (b) Dual-Core scheduling improves FFN throughput by 33.64\%. (c) The combined optimizations yield a total throughput gain of 61.74\%.}
    \label{fig:throughput_ablation}
\end{figure*}

Fig.~\ref{fig:throughput_ablation} isolates the throughput gains from our scheduling strategies:
\begin{itemize}
    \item \textbf{Attention:} Head-Level Pipelining (HLP) hides heterogeneous latency, improving throughput by \textbf{118.87\%}.
    \item \textbf{FFN:} The Dual-Core strategy co-executes the massive ternary weight projections across both the specialized TINT-Cores and the versatile BoothFlex-Core, improving throughput by \textbf{33.64\%}.
\end{itemize}
Overall, the combined optimizations yield a total throughput gain of \textbf{61.74\%}.

\subsubsection{Selection of Top-K and $M_{\text{unified}}$}
\label{sec:top_k_and_max}
\begin{figure*}[t]
    \centering
    \begin{tikzpicture}
        \begin{groupplot}[
            group style={group size=2 by 1, horizontal sep=1.15cm},
            width=0.455\textwidth,
            height=0.305\textwidth,
            xmin=16, xmax=48,
            xtick={16,24,32,40,48},
            xlabel={Top-K},
            ylabel={Longer-context Perplexity},
            grid=both,
            major grid style={dashed,gray!30},
            minor grid style={dotted,gray!20},
            tick label style={font=\footnotesize},
            label style={font=\footnotesize},
            title style={font=\footnotesize},
            legend style={
                font=\footnotesize,
                draw=none,
                at={(0.5,-0.24)},
                anchor=north,
                legend columns=5,
                /tikz/every even column/.append style={column sep=0.25cm}
            },
            every axis plot/.append style={thick, mark=*, mark size=1.8pt},
        ]

        \nextgroupplot[
            title={C4},
            ymin=10.0, ymax=12.2
        ]
        \addplot coordinates {(16,10.49435) (24,10.31132) (32,10.21124) (40,10.14495) (48,10.09345)};
        \addlegendentry{Max=12}

        \addplot coordinates {(16,10.49632) (24,10.31946) (32,10.21975) (40,10.15415) (48,10.10875)};
        \addlegendentry{Max=14}

        \addplot coordinates {(16,10.42930) (24,10.27044) (32,10.18144) (40,10.12468) (48,10.08434)};
        \addlegendentry{Max=16}

        \addplot coordinates {(16,10.69951) (24,10.56820) (32,10.49640) (40,11.44904) (48,11.41612)};
        \addlegendentry{Max=18}

        \addplot coordinates {(16,12.07016) (24,11.96875) (32,11.92738) (40,11.91085) (48,11.90525)};
        \addlegendentry{Max=20}

        \nextgroupplot[
            title={Wiki},
            ymin=10.0, ymax=12.8
        ]
        \addplot coordinates {(16,10.66084) (24,10.42643) (32,10.30089) (40,10.23002) (48,10.18162)};
        \addplot coordinates {(16,10.66898) (24,10.44392) (32,10.32511) (40,10.25398) (48,10.20882)};
        \addplot coordinates {(16,10.64006) (24,10.40947) (32,10.28934) (40,10.21991) (48,10.17331)};
        \addplot coordinates {(16,10.80815) (24,10.58302) (32,10.48138) (40,12.27012) (48,12.25128)};
        \addplot coordinates {(16,12.58057) (24,12.45882) (32,12.41256) (40,12.37638) (48,12.36216)};

        \end{groupplot}
    \end{tikzpicture}
    \caption{\revised{Impact of Top-K and $M_{\text{unified}}$.}}
    \label{fig:topk_maxvalue}
    \vspace{-6pt}
\end{figure*}
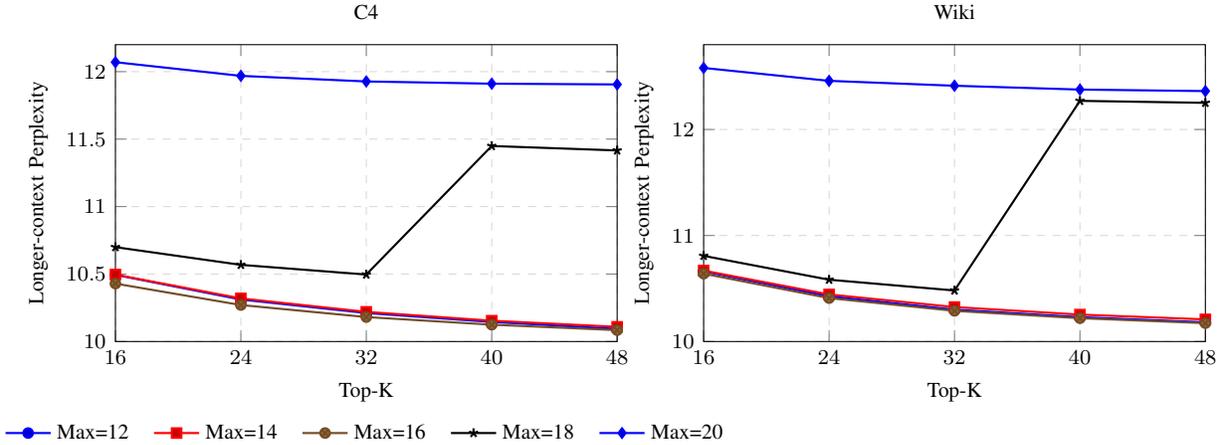

\revised{
The selection of the Top-K and $M_{\text{unified}}$ parameters affects the model performance, latency and hardware cost. For $M_{\text{unified}}$, the selection method in~\cite{flashdecoding++2024} is adopted here, i.e. 99.99\% of the values within the $M_{\text{unified}}$. For the Bitnet 1.58b 3B model, we select $M_{\text{unified}}=16$. For Top-K, in general, larger K values will have lower performance loss due to gradually approaching the original one, but this also leads to higher hardware cost and latency. This paper selects the K value empirically by selecting Top-K tokens with over 95\% of the attention energy. For the Bitnet 1.58b 3B model, K is selected as 32. 

Table~\ref{tab:long_context_ppl} shows evaluates the impact of the proposed LOP approximation under longer-context settings. Using Top-32 LOP and $M_{\text{unified}}=16$, the perplexity degradation remains limited for sequence lengths up to 2048 on both C4 and Wikitext datasets, while a slightly larger gap is observed at 4096. This result suggests that the proposed configuration is robust in practical operating regimes, although extremely long contexts may require additional hyperparameter tuning.

Fig.~\ref{fig:topk_maxvalue} evaluate different combinations of Top-K and $M_{\text{unified}}$ on the C4 and Wiki dataset for the Bitnet 1.58b 3B model. As shown in Fig.~\ref{fig:topk_maxvalue}, increasing Top-K consistently improves perplexity when Max Value is set to 12, 14, or 16. Among all configurations, $M_{\text{unified}}$ = 16  achieves the best overall performance, whereas larger Max Value settings (18 and 20) noticeably degrade accuracy,  suggesting that an overly large value range can destabilize the approximation..

}

\begin{table}[t]
\centering
\caption{\revised{Longer-context perplexity under Top-32 LOP and unified max value $M_{\text{unified}}=16$.}}
\label{tab:long_context_ppl}
\setlength{\tabcolsep}{5pt}
\renewcommand{\arraystretch}{1.1}
\begin{tabular}{c c c c c c c}
\toprule
\multirow{2}{*}{\textbf{Dataset}} & \multirow{2}{*}{\textbf{Setting}} & \multicolumn{5}{c}{\textbf{Sequence Length}} \\
\cmidrule(lr){3-7}
 &  & \textbf{256} & \textbf{512} & \textbf{1024} & \textbf{2048} & \textbf{4096} \\
\midrule
\multirow{2}{*}{C4}
  & w/o LOP & 11.43 & 10.55 & 10.06 &  9.84 & 16.63 \\
  & w/ LOP  & 11.49 & 10.68 & 10.30 & 10.18 & 17.93 \\
\midrule
\multirow{2}{*}{Wikitext-2}
  & w/o LOP & 15.94 & 13.14 & 11.17 &  9.92 & 19.45 \\
  & w/ LOP  & 15.95 & 13.21 & 11.35 & 10.29 & 20.58 \\
\bottomrule
\end{tabular}
\end{table}

\subsection{Comparison with State-of-the-Art Works}
\label{subsec:comparison}

We compare VitaLLM against \textbf{Slim-Llama}~\cite{slimllama2025} (28nm ASIC) and FPGA-based designs \textbf{TerEffic}~\cite{tereffic} and \textbf{TeLLMe}~\cite{tellme2025, tellmev2_2025}. Table~\ref{tab:comparison_sota} summarizes the key performance metrics.\revised{The data of the other designs are extracted from their publications.}

\begin{table*}[t]
    \centering
    \caption{Comparison with state-of-the-art ternary LLM accelerators.}
    \label{tab:comparison_sota}
    \small
    \resizebox{\textwidth}{!}{%
    \begin{tabular}{lccccccc}
        \hline
        \textbf{Metric} & \textbf{TeLLMe v2}~\cite{tellmev2_2025} & \textbf{TeLLMe}~\cite{tellme2025} & \textbf{TerEffic}~\cite{tereffic} & \textbf{Slim-Llama}~\cite{slimllama2025} & \revised{\textbf{TENET}~\cite{tenet}} & \revised{\textbf{TOM}~\cite{tom}} & \textbf{VitaLLM (Ours)} \\ \hline
        \textbf{Platform} & FPGA KV260 & FPGA KV260 & FPGA U280 & ASIC 28nm & ASIC 28nm & ASIC 7nm & \textbf{ASIC 16nm} \\
        \textbf{Frequency (MHz)} & 250 & 250 & 150 & 25-200 & 500 & 500 & \textbf{1000} \\
        \textbf{Voltage (V)} & - & - & - & 0.58-1.0 & - & - & \textbf{0.8} \\
        \textbf{On-chip Mem.} & 98.5\% BRAM & 71\% BRAM & 42 MB & 500 KB & 1.38 MB & $\sim$536 MB & \textbf{130.5 KB} \\
        \textbf{Area ($\text{mm}^2$)} & - & - & - & 20.25 & 91.0 & 56.9 & \textbf{0.223} \\
        \textbf{Power (mW)} & 4800 & 6720 & 46200 & 82.07 & 5700 & 5330 & \textbf{65.97} \\ \hline
        \textbf{Model Params} & BitNet 0.73B & BitNet 0.73B & BitNet 2.7B & BitNet 3B & BitNet 3B & BitNet 2B & BitNet 3B \\
        \textbf{Prefill Time (s)} & 0.45 & 0.55 & - & 0.635 & - & - & 0.88 \\
        \textbf{Throughput (tk/s)} & 24.6 & 9.51 & 727 & - & - & 3306 & \textbf{70.70} \\ \hline
        \textbf{Area Eff. (GOPS/$\text{mm}^2$)} & - & - & - & 242.96 & - & - & \textbf{1147.98} \\
        \textbf{Energy Eff. (tk/J)} & 5.13 & 1.42 & 15.74 & - & - & 620.26 & \textbf{1098.38} \\
        \textbf{FOM (TOPS/$\text{mm}^2$/W)} & - & - & - & 2.96 & - & - & \textbf{17.4} \\ \hline
    \end{tabular}%
    }
\end{table*}

\begin{table}[t]
\centering
\caption{\revised{Inference performance of different model size.}}
\label{tab:bitnet_edge_perf}
\begin{tabular}{lcccc}
\toprule
\textbf{Model} & \textbf{2B} & \textbf{3B} & \textbf{7B} & \textbf{13B} \\
\midrule
Prefill(s) (64 tokens) & 0.65 & 0.88 & 1.76 & 3.44 \\
Throughput (tk/s)                  & 99.21 & 70.70 & 36.46 & 18.62 \\
\bottomrule
\end{tabular}
\end{table}

\subsubsection{Performance Analysis}
\label{sec:performance analysis}
VitaLLM achieves a decoding throughput of \textbf{70.70 tokens/s}, well exceeding the real-time target of 20 tokens/s. Power consumption is reduced to \textbf{65.97 mW}, a 19.6\% improvement over Slim-Llama. While the prefill latency is higher than Slim-Llama's, this is an intentional design trade-off. By avoiding over-provisioning for the compute-bound prefill stage, our dual-core architecture maintains high utilization during the bandwidth-bound decode stage while ensuring valid real-time interactivity ($<1.0$s). \revised{Table~\ref{tab:bitnet_edge_perf} show the throughput of the different model sizes by configuring the hardware through size related parameters.}

\subsubsection{Area and Efficiency}
VitaLLM demonstrates exceptional compactness, occupying only \textbf{0.223 $\text{mm}^2$}. Furthermore, efficient scheduling minimizes on-chip memory to \textbf{130.5 KB}, significantly lower than FPGA baselines. Consequently, VitaLLM achieves a Figure of Merit (FOM) of \textbf{17.4 TOPS/$\text{mm}^2$/W}, validating the efficiency of the proposed hardware-software co-design.

%% file: chapters/5ExtendedDesign.tex
\section{Extended Design: BoothFlex-BS Core}
\label{sec:extended_design}

While VitaLLM demonstrates exceptional efficiency for BitNet b1.58, the rapidly evolving landscape of LLM quantization often requires varying levels of precision (e.g., INT4, INT8, INT16) to balance accuracy and performance across different layers or deployment scenarios. To address this need for versatility, we introduce the \textbf{BoothFlex-BS Core}, a bit-serial and precision-agile extension of our unified engine. By processing data in granular chunks, this core enables the hardware to dynamically support arbitrary integer bit-widths, trading latency for precision without architectural redesign.

\subsection{Bit-Serial Architecture}
\label{subsec:bs_architecture}

To overcome the rigidity of fixed-precision datapaths, the BoothFlex-BS Core adopts a \textbf{Bit-Serial} approach. Unlike the original BoothFlex-Core which processes the entire input vector in a single cycle, this extended core decomposes the input activation into 4-bit segments called ``nibbles'' and processes them sequentially.

\subsubsection{Nibble Slicing and Hierarchical Accumulation}
Mathematically, a multi-bit value $Y$ is processed as a sum of its shifted nibbles: $Y = \sum_{i=0}^{N-1} Y_i \cdot 2^{4i}$. During inference, the core executes a hierarchical accumulation strategy:
\begin{itemize}
    \item \textbf{Temporal Accumulation (Shift-by-4):} Within each PE, activation nibbles are streamed from MSB to LSB. In each cycle, the accumulator is left-shifted by 4 bits before adding the partial product of the new nibble ($Accum \leftarrow (Accum \ll 4) + PP_{nibble}$).
    \item \textbf{Spatial Accumulation (Shift-by-2):} To align with the Radix-4 Booth encoding (which has a step size of 2 bits), results passed between Booth levels are left-shifted by 2 bits.
\end{itemize}
This ``Shift-and-Add'' mechanism allows the same physical 4-bit multiplier hardware to support 4, 8, 12, or 16-bit operations simply by extending the execution cycles.

\subsubsection{Microarchitecture Implementation}
\begin{figure}[t]
    \centering
    \includegraphics[width=0.98\linewidth]{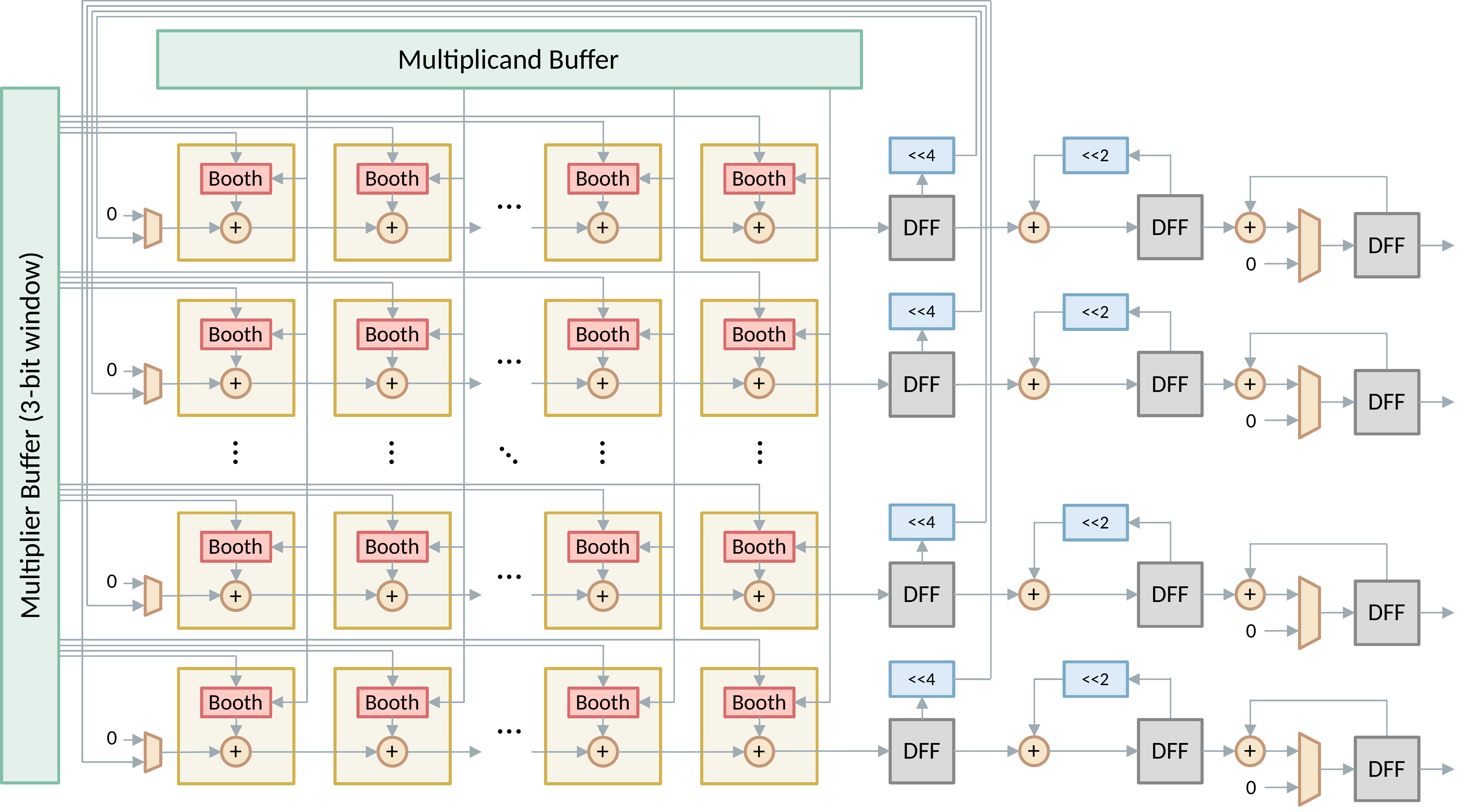}
    \caption{Microarchitecture of the BoothFlex-BS Core. The design integrates a 4-bit Bit-Serial MAC unit with Shift-Add logic to enable iterative precision reconstruction.}
    \label{fig:boothflex_bs_arch}
\end{figure}

Fig.~\ref{fig:boothflex_bs_arch} depicts the microarchitecture. Key modifications include: (1) a \textbf{Multiplicand Buffer} that feeds 4-bit nibbles to the datapath; (2) a \textbf{Bit-Serial MAC Unit} that retains the efficient Radix-4 Booth encoding of the original design but operates on 4-bit chunks; and (3) \textbf{Shift-Add Logic} to reconstruct high-precision results iteratively. This hybrid approach significantly reduces combinational logic depth compared to full-width multipliers while retaining Booth encoding benefits.

\subsection{Comparison and Trade-off Analysis}
\label{subsec:comparison_tradeoff}

\subsubsection{Core-Level Efficiency}
We compare BoothFlex-BS with \textbf{BitMoD}~\cite{bitmod2024}, a state-of-the-art bit-serial accelerator. As shown in Table~\ref{tab:comparison_bitmod}, by optimizing strictly for integer arithmetic (dominant in edge LLMs) rather than mixed FP/INT workloads, our design achieves an extremely compact area and superior energy efficiency.

\begin{table}[h]
    \centering
    \caption{Core-level comparison with BitMoD.}
    \label{tab:comparison_bitmod}
    \small
    \begin{tabular}{lcc}
        \hline
        \textbf{Metric} & \textbf{BitMoD}~\cite{bitmod2024} & \textbf{BoothFlex-BS (Ours)} \\ \hline
        \textbf{Tech / Freq} & 28nm / 1GHz & 16nm / 1GHz \\
        \textbf{Precision} & INT/FP $\times$ FP16 & INT $\times$ INT (Variable) \\ \hline
        \textbf{Area ($\bm{\mathrm{\mu m^2}}$)} & 99,509 & \textbf{5,145} \\
        \textbf{Power (mW)} & 39.36 & \textbf{5.07} \\ \hline
        \textbf{FOM} & 16.34 & \textbf{2,453.5} \\ \hline
    \end{tabular}
\end{table}

\subsubsection{System-Level Trade-offs}
To quantify the cost of flexibility, we evaluate an extended VitaLLM system where all computing cores are replaced by BoothFlex-BS Cores (Table~\ref{tab:comparison_original_extended}).

\begin{table}[h]
    \centering
    \caption{System-level trade-off: Original vs. Extended Design.}
    \label{tab:comparison_original_extended}
    \small
    \begin{tabular}{lcc}
        \hline
        \textbf{Metric} & \textbf{Original} & \textbf{Extended (BS)} \\ \hline
        \textbf{Precision} & Fixed (Ternary$\times$INT8) & \textbf{Flexible (INT$\times$INT)} \\
        \textbf{Area / Power} & Baseline & +4.5\% / +0.6\% \\
        \textbf{Throughput} & \textbf{70.70 tk/s} & 35.38 tk/s \\ \hline
    \end{tabular}
\end{table}

The results reveal a clear trade-off:
\begin{itemize}
    \item \textbf{Minimal Hardware Cost:} The extended design incurs negligible overhead in area (+4.5\%) and power (+0.6\%), proving that the precision-agile logic is lightweight.
    \item \textbf{Throughput Impact:} Decoding speed drops by approximately 50\%. This is expected as the bit-serial core requires 2 cycles to process an 8-bit activation (two 4-bit nibbles), whereas the original TINT-Core completes it in a single cycle.
\end{itemize}

In conclusion, the extended design offers a viable alternative for scenarios prioritizing multi-precision support, maintaining real-time performance with minimal area penalty.

%% file: chapters/6Conclusion.tex
\section{Conclusion}
\label{sec:conclusion}

This paper presents VitaLLM, a hardware-software co-design accelerator tailored for BitNet b1.58 inference on edge devices. To overcome the utilization bottlenecks and memory wall inherent in ternary LLMs, we introduce a heterogeneous Dual-Core Compute Strategy combined with Leading One Prediction (LOP) and a Dependency-Aware Scheduling framework. These innovations effectively decouple computational workloads, prune redundant memory accesses, and hide execution latency, ensuring high efficiency across both prefill and decode stages.

Silicon implementation in TSMC 16nm demonstrates that VitaLLM achieves a decoding throughput of 70.70 tokens/s within an ultra-compact 0.223 $\text{mm}^2$ area, delivering a state-of-the-art Figure of Merit (FOM) of 17.4 TOPS/$\text{mm}^2$/W. Furthermore, the extended BoothFlex-BS design highlights the adaptability of the architecture for precision-agile inference. Overall, this work validates that holistic cross-layer optimization enables high-performance LLM deployment under stringent edge constraints.

%% file: bib/ieeeBSTcontrol.bib
@IEEEtranBSTCTL{IEEEexample:BSTcontrol,
CTLdash_repeated_names= "no",
}


%% file: bib/thesis.bib
@article{bitnet_cpp,
  title={{1-bit AI Infra: Part 1.1, Fast and Lossless BitNet b1.58 Inference on CPUs}},
  author={Wang, Jinheng and Zhou, Hansong and Song, Ting and Mao, Shaoguang and Ma, Shuming and Wang, Hongyu and Xia, Yan and Wei, Furu},
  journal={arXiv preprint arXiv:2410.16144},
  year={2024},
  month={Oct}
}

@article{survey_hw_llm,
  title={{A Survey on Hardware Accelerators for Large Language Models}},
  author={Kachris, Christoforos},
  journal={arXiv preprint arXiv:2401.09890},
  year={2024},
  month={Jan}
}

@inproceedings{mobilellm2024,
  title={{MobileLLM: Optimizing sub-billion parameter language models for on-device use cases}},
  author={Liu, Zechun and Zhao, Changsheng and Xiong, Yunyang and Chang, Ernie and Iandola, Forrest and Lai, Chen and Tian, Yuandong and Fedorov, Igor and Shi, Yangyang and Krishnamoorthi, Raghuraman and others},
  booktitle={Proceedings of the 41st International Conference on Machine Learning (ICML)},
  year={2024},
  month={Jul},
}

@article{energon2022,
  title={{Energon: Toward efficient acceleration of transformers using dynamic sparse attention}},
  author={Zhou, Zhe and Liu, Junlin and Gu, Zhenyu and Sun, Guangyu},
  journal={IEEE Transactions on Computer-Aided Design of Integrated Circuits and Systems (TCAD)},
  volume={42},
  number={1},
  pages={136--149},
  year={2023},
  month={Jan}
}

@inproceedings{fact2023,
  title={{FACT: FFN-attention co-optimized transformer architecture with eager correlation prediction}},
  author={Qin, Yubin and Wang, Yang and Deng, Dazheng and Zhao, Zhiren and Yang, Xiaolong and Liu, Leibo and Wei, Shaojun and Hu, Yang and Yin, Shouyi},
  booktitle={Proceedings of the 50th Annual International Symposium on Computer Architecture (ISCA)},
  pages={1--14},
  year={2023},
  month={Jun}
}

@inproceedings{slimllama2025,
  title={{Slim-Llama: A 4.69mW large-language-model processor with binary/ternary weights for billion-parameter Llama model}},
  author={Kim, Sangyeob and Lee, Jungwan and Yoo, Hoi-Jun},
  booktitle={IEEE International Solid-State Circuits Conference (ISSCC)},
  pages={422--422},
  year={2025},
  month={Feb}
}

@inproceedings{flashdecoding++2024,
  title={{FlashDecoding++: Faster large language model inference with asynchronization, flat GEMM optimization, and heuristics}},
  author={Hong, Ke and Dai, Guohao and Xu, Jiaming and Mao, Qiuli and Li, Xiuhong and Liu, Jun and Chen, Kangdi and Dong, Yuhan and Wang, Yu},
  booktitle={Proceedings of the 7th Conference on Machine Learning and Systems (MLSys)},
  year={2024}
}

@inproceedings{attacc2024,
  title={{AttAcc! Unleashing the power of PIM for batched transformer-based generative model inference}},
  author={Park, Jaehyun and Choi, Jaewan and Kyung, Kwanhee and Kim, Michael Jaemin and Kwon, Yongsuk and Kim, Nam Sung and Ahn, Jung Ho},
  booktitle={Proceedings of the 29th ACM International Conference on Architectural Support for Programming Languages and Operating Systems (ASPLOS)},
  volume={2},
  pages={103--119},
  year={2024},
  month={Apr}
}

@inproceedings{bitmod2024,
  title={{BitMoD: Bit-serial mixture-of-datatype LLM acceleration}},
  author={Chen, Yuzong and AbouElhamayed, Ahmed F. and Dai, Xilai and Wang, Yang and Andronic, Marta and Constantinides, George A. and Abdelfattah, Mohamed S.},
  booktitle={Proceedings of the 31st IEEE International Symposium on High-Performance Computer Architecture (HPCA)},
  year={2025},
  month={Mar}
}

@article{bitnet158,
  title={{The era of 1-bit LLMs: All large language models are in 1.58 bits}},
  author={Ma, Shuming and Wang, Hongyu and Ma, Lingxiao and Wang, Lei and Wang, Wenhui and Huang, Shaohan and Dong, Li and Wang, Ruiping and Xue, Jilong and Wei, Furu},
  journal={arXiv preprint arXiv:2402.17764},
  year={2024},
  month={Feb}
}

@article{sofa2024,
  title={{SOFA: A compute-memory optimized sparsity accelerator via cross-stage coordinated tiling}},
  author={Wang, Huizheng and Fang, Jiahao and Tang, Xinru and Yue, Zhiheng and Li, Jinxi and Qin, Yubin and Guan, Sihan and Yang, Qize and Wang, Yang and Li, Chao and others},
  journal={arXiv preprint arXiv:2407.10416},
  year={2024},
  month={Jul}
}

@article{bitnet2b4t,
  title={{BitNet b1.58 2B4T technical report}},
  author={Ma, Shuming and Wang, Hongyu and Huang, Shaohan and Zhang, Xingxing and Hu, Ying and Song, Ting and Xia, Yan and Wei, Furu},
  journal={arXiv preprint arXiv:2504.12285},
  year={2025},
  month={Apr}
}

@article{tereffic,
  title={{TerEffic: Highly efficient ternary LLM inference on FPGA}},
  author={Yin, Chenyang and Bai, Zhenyu and Venkatram, Pranav and Aggarval, Shivam and Li, Zhaoying and Mitra, Tulika},
  journal={arXiv preprint arXiv:2502.16473},
  year={2025},
  month={May}
}

@article{tellme2025,
  title={{TeLLMe: An energy-efficient ternary LLM accelerator for prefill and decode on edge FPGAs}},
  author={Qiao, Ye and Chen, Zhiheng and Zhang, Yifan and Wang, Yian and Huang, Sitao},
  journal={arXiv preprint arXiv:2504.16266},
  year={2025},
  month={Apr}
}

@article{tellmev2_2025,
  title={{TeLLMe v2: An efficient end-to-end ternary LLM prefill and decode accelerator with table-lookup matmul on edge FPGAs}},
  author={Qiao, Ye and Chen, Zhiheng and Zhang, Yifan and Wang, Yian and Huang, Sitao},
  journal={arXiv preprint arXiv:2510.15926},
  year={2025},
  month={Oct}
}

@inproceedings{liang2024computing,
  title={{Computing architecture for large language models (LLMs) and large multimodal models (LMMs)}},
  author={Liang, Bor-Sung},
  booktitle={Proceedings of the 2024 International Symposium on Physical Design (ISPD)},
  year={2024}
}

@article{saha2023kdegree,
  title={{k-degree parallel comparison-free hardware sorter for complete sorting}},
  author={Ray, Sanchita Saha and Ghosh, Surajeet},
  journal={IEEE Transactions on Computer-Aided Design of Integrated Circuits and Systems (TCAD)},
  volume={42},
  number={5},
  pages={1438--1449},
  year={2023},
  month={May}
}

@article{tom,
  title={TOM: A Ternary Read-only Memory Accelerator for LLM-powered Edge Intelligence},
  author={Guan, Hongyi and Zhang, Yijia and Wang, Wenqiang and Gao, Yizhao and Cao, Shijie and Zhang, Chen and Xu, Ningyi},
  journal={arXiv preprint arXiv:2602.20662},
  year={2026}
}

@article{tenet,
  title={Tenet: An efficient sparsity-aware lut-centric architecture for ternary llm inference on edge},
  author={Huang, Zhirui and Ma, Rui and Cao, Shijie and Shu, Ran and Wang, Ian and Cao, Ting and Chen, Chixiao and Xiong, Yongqiang},
  journal={arXiv preprint arXiv:2509.13765},
  year={2025}
}
